\documentclass{JHEP3}
\usepackage[dvips]{graphicx}
\usepackage{amsmath,amssymb}

\newcommand{\be}{\begin{equation}}
\newcommand{\ee}{\end{equation}}
\newcommand{\bea}{\begin{eqnarray}}
\newcommand{\eea}{\end{eqnarray}}
\newcommand{\fatk}{\mathbf k}

\newcommand{\fate}{\mathbf e}
\newcommand{\mat}{\left ( \begin{array}{cc}}
\newcommand{\emat}{\end{array} \right )}

\newcommand{\bfe}{{\mathbf e}}
\newcommand{\bfk}{{\mathbf k}}
\newcommand{\bfl}{{\mathbf l}}
\newcommand{\bfm}{{\mathbf m}}
\newcommand{\bfn}{{\mathbf n}}
\newcommand{\bfq}{{\mathbf q}}
\newcommand{\bfr}{{\mathbf r}}
\newcommand{\bfx}{{\mathbf x}}

\newcommand{\bz}{\overline{z}}

\newcommand{\bpsi}{\overline{\psi}}

\newcommand{\hmu}{\hat{\mu}}

\newcommand{\cN}{{\cal N}}

\newcommand{\Z}{\mathbb{Z}}
\newcommand{\R}{\mathbb{R}}

\newcommand{\Q}{\mathbb{Q}}

\newcommand{\nn}{\nonumber}

\newcommand{\Tr}{{\rm Tr}\,}
\newcommand{\del}{\partial}
\newcommand{\e}{\epsilon}

\title{Classification of Supersymmetric Lattice Gauge Theories by Orbifolding}
\author{Poul H. Damgaard and So Matsuura\\ The Niels Bohr Institute, 
The Niels Bohr International Academy,
Blegdamsvej 17, DK-2100 Copenhagen, Denmark}
\abstract{We provide a general classification of supersymmetric lattice gauge
theories that can be obtained from orbifolding of theories with four and eight
supercharges.
We impose at least one preserved supercharge on the lattice and Lorentz
invariance in the naive continuum limit. 
Starting with four supercharges, 
we obtain one two-dimensional lattice gauge theory, 
identical to the one already given in the literature. 
Starting with eight supercharges, 
we obtain a unique three-dimensional lattice gauge theory 
and infinitely many two-dimensional lattice theories. 
They can be classified according to seven distinct groups, 
five of which have two preserved supercharges while the others have
only one. 
}
\begin{document}

\section{Introduction}

Recently, there has been substantial progress in the formulation of euclidean
lattice gauge theories with remnants of space-time supersymmetry 
\cite{Kaplan:2002wv}--\nocite{Cohen:2003xe,Cohen:2003qw}%
\nocite{Kaplan:2005ta,Endres:2006ic}%
\nocite{Catterall:2001fr,Catterall:2001wx}%
\nocite{Catterall:2003uf,Catterall:2003wd}%
\nocite{Catterall:2005fd,Catterall:2006jw}%
%
%
\nocite{D'Adda:2004jb,D'Adda:2005zk}%
\nocite{Bruckmann:2006ub,Bruckmann:2006kb}%
\nocite{Sugino:2003yb,Sugino:2004qd}%
\nocite{Sugino:2004uv,Sugino:2006uf}%
\nocite{Unsal:2005yh}\cite{Unsal:2006qp}. 
A common feature of almost all of these
new approaches has been the connection to topological field theory through
twisting. The remnants of supersymmetry that are preserved on the lattice 
do not generate Poincar\'{e} invariance and are thus not in conflict with
the reduced lattice group of space-time symmetries. Rather, the generators are nilpotent
operators, completely analogous to the BRST and anti-BRST operators of the
corresponding topological field theories in the continuum. Only these BRST 
and/or anti-BRST symmetries are preserved on the lattice. However, 
if the
naive continuum limit yields the usual twisted formulations of the supersymmetric
theories in question it is hoped that the BRST/anti-BRST symmetries are sufficient
to guarantee that this occurs at the full quantum level of the lattice theories as well.

A systematic approach to these new formulations of supersymmetric lattice
gauge theories is based on the orbifolding technique 
\cite{Kaplan:2002wv}--\cite{Endres:2006ic}.
The idea is, roughly, to
start with a huge gauge group, say $U(kN^d)$ in a 
``mother theory'' that is dimensionally reduced 
to zero dimensions. With no space-time coordinates present all fields are really just
matrix variables living in the adjoint representation of the gauge group.
The lattice itself is now generated out of these matrices by means of an orbifold
projection followed by deconstruction \cite{Arkani-Hamed:2001ca}, 
that is, shifts of the fields. These shifts introduce a basic lattice spacing
$a$ in dimensionful units. The aim is typically a $d$-dimensional lattice theory in a finite
volume $N^d$ and with gauge group $U(k)$. A scheme for doing this on lattices with a 
continuous time variable was first presented in
ref. \cite{Kaplan:2002wv}. 
A few years ago,
Cohen, Kaplan, Katz and Unsal 
\cite{Cohen:2003xe}--\cite{Kaplan:2005ta}
showed how to extent this procedure to the euclidean formulation. 
For a detailed analysis of the orbifold construction of 
supersymmetric lattice theories, 
see \cite{Giedt:2003xr}--\nocite{Giedt:2003ve,Onogi:2005cz}%
\cite{Ohta:2006qz} (see also \cite{Fukaya:2006mg}). 
For a very nice review 
of the orbifold construction of lattice theories we refer to
\cite{Giedt:2006pd}.

Because supersymmetry requires a careful balance of bosonic and fermionic degrees
of freedom, it is evident that lattice prescriptions of such theories must somehow
get around the usual fermion doubling problem. This must hold both at
finite lattice spacing $a$ and in the continuum limit. As stressed in refs.
\cite{Catterall:2001fr}--\cite{Catterall:2006jw}
and \cite{D'Adda:2004jb}--\cite{Bruckmann:2006kb}, 
one clue seems to lie in a underlying connection
to the Dirac-K\"{a}hler formulation \cite{Rabin:1981qj,Becher:1982ud}\footnote{For 
two-dimensional $\cN=(2,2)$ supersymmetric gauge theory, 
Suzuki and Taniguchi have given a lattice formulation without this connection 
\cite{Suzuki:2005dx}, arguing that because of super-renormalizability only
a single one-loop counterterm needs to be adjusted in order to regain supersymmetry.}. 
Alternatively, the twisting that
turns ordinary gauge field theories in the continuum into topological
field theories in the continuum requires a departure from the assignment of
spin and statistics which is imposed by the spin-statistics theorem. By an
``untwisting'' in the continuum the usual multiplet of fields that is in
accordance with the spin-statistics theorem is recovered. This untwisting
is required also for extracting observables from the corresponding lattice
theories. 
 
Related to the delicate balance of fermionic and bosonic degrees of freedom
is the obvious difficulty of reconciling the conventionally used compact
gauge link variables on the lattice with the fermionic partners. In a sense, supersymmetry
balances the ``zero'' of fermionic integrations with an ``infinity'' coming
from non-compact bosonic integrations. 
In Sugino's approach \cite{Sugino:2003yb}--\cite{Sugino:2006uf}, which has
compact gauge variables,
the supersymmetry transformations are modified on the lattice, even in the
case of the topological, nilpotent, symmetries. 
On the orbifolded lattices,
gauge field variables are simply non-compact from the outset, 
and thus unusual from the
lattice perspective. Such an assignment is however perfectly  natural if one sees
the (always non-compact) scalar fields as dimensionally reduced components
of gauge potentials. 

Our aim in this paper is to systematically explore the supersymmetric lattice theories that
can be generated by orbifold projections. In doing so we shall also provide the answers
to the following questions:

\begin{itemize}
\item Given a mother theory with a given number of supercharges, how many different
lattice theories can be generated at fixed space-time dimension $d$ and fixed number
of scalar supercharges on those given lattices?
\item Which of those orbifolded lattices lead to Lorentz invariant theories in the
naive continuum limit?
\item How does the lattice theory depend on the number of field variables that
are shifted after the initial orbifold projection?

\end{itemize}

Remarkably, the classification of supersymmetric lattice theories based on orbifold 
projections turns out to be relatively simple. Some of the supersymmetric lattice gauge
theories that can be generated by this technique have already been described in the 
literature, but not all. In this paper we shall provide what we believe is the complete 
classification of orbifolded
lattice theories with four and eight supercharges. The classification of theories
with sixteen supercharges is a bit more involved, and will be presented in a 
separate publication \cite{DM}.

\section{Target theories with four supercharges}
\label{Target theories with four supercharges}

We will begin by briefly recalling the main ingredients in the construction. The
starting point is a mother theory which lives in zero space-time dimensions. For a
theory with four supercharges we can obtain it by dimensional reduction of
${\cal N}$=1 supersymmetric Yang-Mills theory in four euclidean
dimensions. 
As in \cite{Cohen:2003xe}, 
we take the gauge group to be $U(kN^2)$, in anticipation of at most
two-dimensional lattices in this case. The restriction to $U(kN^2)$ rather than
$SU(kN^2)$ (or other gauge groups) is not essential. After dimensional reduction
the mother theory takes the form
\be
S_{\rm m} = \frac{1}{g^2}{\rm Tr}\left(-\frac{1}{4}[v_{\alpha},v_{\beta}]^2 
+ \frac{i}{2}
\bar{\Psi}\Gamma_\alpha[v_\alpha,\Psi]\right), 
\qquad (\alpha,\beta=0,\cdots,3)
\label{Smother0}
\ee
where $\Gamma_\alpha$ are $SO(4)$ Dirac matrices, 
$v_\alpha$ are $kN^2\times kN^2$ hermitian matrices, 
$\Psi$ is a four-component fermion and $\bar\Psi\equiv \Psi^T C$ with 
the charge conjugation matrix $C$ satisfying, 
\be
C^{-1}\Gamma_{\alpha}C ~=~ -\Gamma_{\alpha}^T ~.
\ee
Following \cite{Cohen:2003xe}, 
we choose a chiral representation of the $\gamma$-matrices,
\be
\Gamma_{\alpha} ~=~ \mat 0  & \sigma_{\alpha} \\
     \bar{\sigma}_{\alpha} &  0 \emat
\ee
with $\sigma_{\alpha} = ({\mathbf 1}, -i\tau_i)$ and $\bar{\sigma}_{\alpha} =
({\mathbf 1},i\tau_i)$. the charge conjugation matrix 
is then represented as 
\be
C ~=~ \Gamma_0\Gamma_2 ~=~ \mat i\tau_2 & 0 \\
                                0 & -i\tau_2 \emat
\ee
It is convenient to decompose the four-spinors into the two-component chiral
components as follows:
\be
\Psi ~\equiv~ {\Psi^{(1)} \choose \Psi^{(2)}} ~~~,~~~~~~ 
\Psi^{(1)} ~\equiv~ {\chi_{12} \choose \eta} ~~~,~~~~~~ 
\Psi^{(2)} ~\equiv~ 
{\psi_2 \choose \psi_1} ~.
\ee
Introducing the complex combinations
\bea
z_1 &\equiv & v_1 + iv_2, \cr
z_2 &\equiv & v_0 + iv_3, 
\eea
the action (\ref{Smother0}) takes the form
\be
S_{\rm m} ~=~ \frac{1}{g^2}{\rm Tr}\left(\frac{1}{4}|[z_m,z_n]|^2 
+ \frac{1}{8}[z_m,\bar{z}_m]^2 + \psi_m[\bar{z}_m,\eta] 
-\chi_{mn}[z_m,\psi_n]\right)
\ee
with $m,n = 1,2$ and $\chi_{mn} = -\chi_{nm}$.

The next step is to identify the maximal number of $U(1)$-symmetries. Because the
mother theory is obtained from a four-dimensional euclidean field theory, it
has inherited the associated $SO(4)$ Lorentz symmetry. In addition, the fermionic 
part of the action is invariant under $U(1)$ chiral rotations. The maximal set of
$U(1)$-symmetries is therefore $U(1)^3$, and we can choose them as 
$SO(2)_{12}\times SO(2)_{03}\times U(1)$,
where the indices refer to the corresponding planes of the original four-dimensional
theory. We denote the abelian charges associated with these three $U(1)$
symmetries by $q_1, q_2$ and $q_3$, respectively. To identify the charges of individual
fermionic components one notes that the 
generators of the original $SO(4)$ rotation symmetries are given by the
commutator, 
\be
\Gamma_{\alpha\beta} ~=~ \frac{i}{4}(\Gamma_\alpha\Gamma_\beta - \Gamma_\beta\Gamma_\alpha) ~=~
\mat \sigma_{\alpha\beta} & 0 \\
      0 & \bar{\sigma}_{\alpha\beta} \emat
\ee
where
\be
\sigma_{\alpha\beta} ~=~ \frac{i}{4}(\sigma_\alpha\bar{\sigma}_\beta - 
\sigma_\beta\bar{\sigma}_\alpha) ~~,~~~~
\bar{\sigma}_{\alpha\beta} ~=~ \frac{i}{4}(\bar{\sigma}_\alpha\sigma_\beta 
- \bar{\sigma}_\beta\sigma_\alpha) . 
\ee
The generator of rotations in the $12$-plane is thus
\be
\Gamma_{12} ~=~ \mat -\frac{1}{2}\tau_3 & 0 \\
                         0 & -\frac{1}{2}\tau_3 \emat
\ee
while the generator of rotations in the $03$-plane is
\be
\Gamma_{03} ~=~ \mat -\frac{1}{2}\tau_3 & 0 \\
                        0 & \frac{1}{2}\tau_3 \emat. 
\ee
We can now fill in the table of $U(1)$ charges. The original symmetry on the
fermions $\Psi$ and $\bar{\Psi}$ corresponds to equal charges $q_3=+1/2$ for
the left chiral components $\chi_{12}$ and $\eta$, and $q_3=-1/2$ for the right  
chiral components $\psi_m$. The two other $U(1)$ charges follow by acting
with the generators shown above. Supplemented with the corresponding $SO(2)$
charges for the complex vector fields $z_m$ and $\bar{z}_m$ this leads to
the charge assignments of Table \ref{charge table 4 SUSY}. 

\begin{table}[h]
\caption{The charge assignment of the maximal $U(1)$ symmetries} 
\begin{center}
\begin{tabular}{c|cccccc}
 & $z_1$ & $z_2$ & $\eta$ & $\chi_{12}$ & $\psi_1$ & $\psi_2$ \\
\hline
$q_1$ & 1 & 0 & 1/2 & -1/2 & -1/2 & 1/2 \\
$q_2$ & 0 & 1 & 1/2 & -1/2 & 1/2 & -1/2 \\
$q_3$ & 0 & 0 & 1/2 & 1/2 & -1/2 & -1/2 \\
\end{tabular}
\end{center}
\label{charge table 4 SUSY}
\end{table}

The mother theory has four supercharges. To ensure at least one unbroken 
supersymmetry in the orbifolded theory we need at least one fermion that
transforms as a singlet under the $U(1)$ symmetries. This may be
intuitively clear from the fact that we precisely wish to keep
those supersymmetry charges that will transform trivially under the reduced
set of Poincar\'{e} symmetries compatible with the generated lattice.
A more direct argument has been given in ref. \cite{Cohen:2003xe}.
We choose the singlet fermion
to be $\eta$. As can be seen from Table 1, the $\eta$ is unique in having
all $q_i$'s equal, while the three other fermions have two $q_i$'s of
-1/2 and one $q_i$ of +1/2. One might therefore expect two classes of
lattice theories, depending on whether $\eta$ or one of $\chi_{12}, \psi_m$
is taken to be a scalar under these $U(1)$ symmetries. 
However, one can easily show that this is not the case. 
Even if we choose another fermion ($\psi_1$, for example) to be a
singlet, we obtain exactly the same orbifolded 
action after a renaming of fields. 
Thus the resulting supersymmetric theory on the abstract orbifolded
lattice is actually unique. 

Because of the constraint that the $\eta$ must have zero charge, we are left 
with two free $U(1)$ symmetries under which all fields should have integer
charges. 
In contrast to previous work \cite{Cohen:2003xe}--\cite{Endres:2006ic},  
we do not insist that these integers be $\pm 1$ 
since our purpose is to construct all possible lattice formulations 
based on orbifolding. 
As there are just two $U(1)$ charges free after fixing the $\eta$ to have
zero charge, we can generate at most two-dimensional lattices in the present case. 
Let us define two charge combinations, 
\bea
r_1 &~\equiv~ & \ell_1^1q_1 + \ell_1^2q_2 - (\ell_1^1+\ell_1^2)q_3, \cr
r_2 &~\equiv~ & \ell_2^1q_1 + \ell_2^2q_2 - (\ell_2^1+\ell_2^2)q_3,
\label{integer U(1) charges 4 SUSY}
\eea
for which $\eta$ automatically has vanishing charge. It is then convenient
to introduce two vectors, 
\be
{\mathbf e}_1 ~\equiv~ {\ell_1^1 \choose \ell_2^1}~~~,~~~~
{\mathbf e}_2 ~\equiv~ {\ell_1^2 \choose \ell_2^2}
\ee
so that the charge assignments of Table \ref{charge table 4 SUSY} 
generalize to the simple form given in Table \ref{remaining charge table 4 SUSY}.
Here, since we are interested in obtaining at least a two-dimensional theory, 
we assume that $\mathbf{e}_1$ and $\mathbf{e}_2$ are linearly independent.
In other words, we can uniquely express any two-dimensional 
vector $\bfk\in\Z^2$ as 
\begin{equation}
 \bfk = \sum_{m=1,2} k_m \bfe_m. \qquad (k_m \in \Z)
 \label{position 4 SUSY}
\end{equation}

\begin{table}[h]
\caption{The two remaining $U(1)$ charges}
\begin{center}
\begin{tabular}{c|cccccc}
 & $z_1$ & $z_2$ & $\eta$ & $\chi_{12}$ & $\psi_1$ & $\psi_2$ \\
\hline
${\mathbf r}$ & ${\mathbf e}_1$ & ${\mathbf e}_2$ & ${\mathbf 0}$ & 
-${\mathbf e}_1$-${\mathbf e}_2$ & ${\mathbf e}_1$ & ${\mathbf e}_2$ 
\end{tabular}
\end{center}
\label{remaining charge table 4 SUSY}
\end{table}

Based on these two remaining $U(1)$ symmetries we can 
now carry out the orbifold
projection. As explained in detail in 
refs. \cite{Cohen:2003xe,Giedt:2006pd} this makes
use of a $Z_N\times  Z_N$ subgroup of $U(1)\times U(1)$. One projects out
all field components that are not rendered invariant by the action of this
$Z_N\times Z_N$ symmetry. In Appendix A we collect some useful formulas
for performing this projection. It can be summarized by an expansion
of all fields in terms of variables living on an abstract lattice
labelled by two-vectors ${\mathbf k}$:
\bea
z_m &~=~& \sum_{\fatk}z_m(\fatk)\otimes E_{\fatk,\fatk+\fate_m} \cr
\bar{z}_m   &~=~& \sum_{\fatk}\bar{z}_m(\fatk)\otimes E_{\fatk+\fate_m,\fatk} \cr
\eta  &~=~& \sum_{\fatk}\eta(\fatk)\otimes E_{\fatk,\fatk} \cr
\psi_m &~=~& \sum_{\fatk}\psi_m(\fatk)\otimes E_{\fatk,\fatk+\fate_m} \cr
\chi_{12} &~=~& \sum_{\fatk}\chi_{12}(\fatk)\otimes
E_{\fatk+\fate_1+\fate_2,\fatk}
\eea
where $z_m(\fatk)$, $\bar{z}_m(\fatk)$ and so on are $k\times k$ matrices 
and $E_{\bfk,\bfl}$ is defined by (\ref{basis matrices}) in Appendix A. 
Making use of the orthogonality relation (\ref{orthogonality relation 2}) 
for the $E$'s, we arrive at an
orbifolded theory described by the abstract lattice action, 
\bea
S_{\rm orb} &=& 
\frac{1}{g^2}{\rm Tr}\sum_{\mathbf k}\Biggl( 
\frac{1}{4}\Bigl|z_m({\mathbf k})
z_n(\fatk+\fate_m) - z_n(\fatk)z_m(\fatk+\fate_n)\Bigr|^2  \cr
&& + \frac{1}{8}\Bigl(z_m(\fatk)\bar{z}_m(\fatk)-\bar{z}_m(\fatk-\fate_m)z_m(\fatk
-\fate_m)\Bigr)^2 \cr
&& + \psi_m(\fatk)\Bigl(\bar{z}_m(\fatk)\eta(\fatk)-\eta(\fatk+\fate_m)
\bar{z}_m(\fatk)\Bigr)\cr 
&& - \frac{1}{2}\chi_{mn}(\fatk)\Bigl(z_m(\fatk)\psi_n(\fatk+\fate_n)
-\psi_n(\fatk)z_m(\fatk+\fate_n)
-(m\leftrightarrow n)\Bigr)\Biggr) ~, \label{Sorb0}
\eea
where we implicitly sum over repeated indices $m,n=1,2$. 
The lattice is periodic and of size
$N\times N$. Variables $z_m(\fatk), 
\bar{z}_m(\fatk), \psi_m(\fatk)$ transform as bifundamentals of $U(k)$,
\be
z_m(\fatk) ~\to~ V(\fatk)^{\dagger}z_m(\fatk)V(\fatk+\fate_m) ~~,~~~
\bar{z}_m(\fatk) ~\to~ V(\fatk+\fate_m)^{\dagger}
\bar{z}_m(\fatk)V(\fatk)~,~~~~~{\rm etc.}
\ee
while $\eta(\fatk)$ transforms as an adjoint under $U(k)$,
\be
\eta(\fatk) ~\to~ V(\fatk)^{\dagger}\eta(\fatk)V(\fatk), 
\ee
and finally $\chi_{12}$ also transforms as a bifundamental,  
\be
\chi_{12}(\fatk) ~\to~ 
V(\fatk + \fate_1 + \fate_2)^\dagger
\chi_{12}(\fatk)
V(\fatk).
\ee
While these transformation rules are similar to those of lattice fields living
on sites, links, and corners (or, alternatively, diagonal links), 
there is yet no space-time lattice, no lattice
spacing $a$, and no kinetic energy terms ``hopping'' between different sites.
Just as the mother theory can be viewed as Eguchi-Kawai large-$N$ reduction 
\cite{Eguchi:1982nm} in the continuum, orbifolding is reminiscent of the
similar Eguchi-Kawai
reduction in a finite volume \cite{Orland:1983gt}. 

We can now see why we only those supersymmetry charges that have
vanishing $U(1)$-charges will be preserved on the lattice. In the
mother theory we have an exact Leibniz rule for the way supersymmetry charges
act on products of fields.
Consider now the way a supersymmetry charge will act on the lattice
variables...

As shown in ref. \cite{Cohen:2003xe}, shifts of the variables
$z_m$ (and $\bar{z}_m$) generate kinetic energy terms for all fields. 
This is a consequence of the $U(k)$ symmetry which automatically induces
covariant derivatives once kinetic terms are introduced for the $z_m$
variables. 
In general, one can shift $N_{\rm shift}$ variables $z_m$, and if we want the target space
theory to be Lorentz invariant in $d$ dimensions it is clear that we 
need $N_{\rm shift} \geq d$. In the present case there is
not much room left (we are not interested in one-dimensional theories), and we choose 
$N_{\rm shift}=d=2$. Then, since $z_m$ has classical dimension one,
\be
z_m(\fatk) ~\to~ \frac{1}{a_m} + z_m(\fatk), 
\label{zshifts0}
\ee
for $m=1,2$. Here the $a_m$'s have the dimension of length. 
After the shifts (\ref{zshifts0}) the action (\ref{Sorb0}) 
takes the following form: 
\bea
S_{\rm lat}^{d=2,N=2} &=& \frac{1}{g^2}{\rm Tr}\sum_{\mathbf k}
\Biggl(\frac{1}{4}\Bigl|
\nabla_m^+ z_n(\bfk) - \nabla_n^+ z_m(\bfk) 
+z_m({\mathbf k})
z_n(\fatk+\bfe_m) - z_n(\fatk)z_m(\fatk+\bfe_n)\Bigr|^2  \cr
&& + \frac{1}{8}\Bigl(
\nabla_m^{+}\Bigl(z_m(\bfk)+\bz_m(\bfk)\Bigr)
+z_m(\fatk+\bfe_m)\bar{z}_m(\bfk+\bfe_m)
-\bar{z}_m(\fatk)z_m(\fatk)\Bigr)^2 \cr
&& + \psi_m(\fatk)\Bigl(
\nabla_m^+ \eta(\bfk)
-\bar{z}_m(\fatk)\eta(\fatk)+\eta(\fatk+\bfe_m)
\bar{z}_m(\fatk)\Bigr)\cr 
&& + \frac{1}{2}\chi_{mn}(\fatk)\Bigl(
\nabla_m^+ \psi_n(\bfk)
+z_m(\fatk)\psi_n(\fatk+\bfe_m)
-\psi_n(\fatk)z_m(\fatk+\bfe_n)
-(m\! \leftrightarrow\! n)\Bigr)\Biggr) ~,\cr
&&
\label{lattice action 4 SUSY}
\eea
where, for an arbitrary function $\phi$, we have introduced the forward 
difference, 
\bea
\nabla_m^+\phi(\fatk) &~=~& \frac{1}{a_m}\left(\phi(\fatk + \bfe_m) -
\phi(\fatk)\right).
\label{difference operation}
\eea
Note that the $a_m$'s can take arbitrary complex values; 
$a_m\equiv |a_m|e^{ib_m}$ in general. 
However, the phase factors $e^{ib_m}$ can be absorbed by proper 
$U(1)$ rotations for the fields. 
Thus we can assume $a_m\in \R_+$. 

It is remarkable that the kinetic terms in 
(\ref{lattice action 4 SUSY}) are defined between nearest neighbors, 
even though we have not restricted the $U(1)$ charges 
(\ref{integer U(1) charges 4 SUSY}) to be just $\pm 1$. 
This is a direct consequence of the deconstruction that is used
to create the kinetic terms. If we fix the values of $a_m$'s, 
the action (\ref{lattice action 4 SUSY}) describes the same lattice 
theory even if we change $\bfe_m$'s 
as long as $\bfe_1$ and $\bfe_2$ are linearly independent. 
The lattice theory obtained from the orbifolding
procedure is thus uniquely labelled by 1) the values of $a_m$'s and 2) 
any linear relation among $\bfe_m$'s. 
In fact, the arguments of the fields 
in this theory can be labelled by a set of
integers $\{k_m\}$ as (\ref{position 4 SUSY}), which is invariant
under a change of basis $\{\bfe_m\}$.

Next, we must consider the naive continuum limit. 
To this end, it is convenient to introduce an invertible linear mapping, 
\be 
f:~~ \fate_m \mapsto \gamma_m a_m\hat{\mu}_m, 
\label{emumap}
\ee
where $\gamma_m$ are arbitrary positive numbers and 
$\hat{\mu}_m$'s are unit vectors. This map gives a one-to-one
association of a space-time lattice with the abstract lattice
space generated by the set ${\{\bfe_m\}}$. 
In particular, the lattice spacings are given by this mapping.  
An abstract vector $\bfk$ is thus mapped naturally to a space-time 
position through  
\begin{equation}
 f: \bfk \mapsto \sum_{m=1,2}k_m \gamma_m a_m \hmu_m.  
\end{equation}
As we shall see, we get a non-trivial restriction on this
map by insisting on Lorentz invariance 
in the naive continuum limit.
In order to make our notation simple, 
we will often, unless there is an obvious ambiguity, 
use the same notation $\bfk$ to
express the space-time position in the following.
The continuum limit is thus defined by
\footnote{
More precisely, we first introduce a common parameter $a$ to set a scale
and write $a_m\equiv a\beta_m$, with $\beta_m$ fixed.
The continuum limit is then defined by $a\to 0$. 
} 
\begin{equation}
 a_m ~\to~ 0, 
\label{define continuum limit}
\end{equation}
so that the difference (\ref{difference operation}) 
becomes the derivative in the continuum
limit, {\em viz.}, 
\begin{equation}
 \frac{1}{a_m}\left(
\phi(\bfk+\gamma_m a_m\hmu_m)-\phi(\bfk) \right)
\to \gamma_m \hmu_m\cdot \vec{\del} \phi(\bfk). 
\label{continuum limit of difference operation}
\end{equation}

Finally, we must find a proper set of space-time basis 
vectors $\{\hmu_m\}$ and values of $\gamma_m$'s for which 
the continuum theory is Lorentz invariant.  
It is sufficient to look at the kinetic term of the bosonic fields, 
\begin{equation}
 \frac{1}{4}\Bigl|\nabla_m^+ z_n(\bfk) - \nabla_n^+ z_m(\bfk) \Bigr|^2
 +\frac{1}{8}\Bigl(\nabla_m^+\left(z_m(\bfk)+\bz_m(\bfk)\right)\Bigr)^2. 
\label{kinetic term 1}
\end{equation}
By expressing $z_m(\bfk)$ and $\bz_m(\bfk)$ as 
\begin{align}
 z_m(\bfk) &\equiv S_m(\bfk) + i T_m(\bfk), \nn \\
 \bz_m(\bfk) &\equiv  S_m(\bfk) - i T_m(\bfk),
\end{align}
the continuum limit of (\ref{kinetic term 1}) can be written as 
\begin{equation}
 -S_m(\bfx)\Bigl(\gamma_l \hmu_l\cdot\vec\del\Bigr)^2 S_m(\bfx)
 -T_m(\bfx)\Bigl[\left(\gamma_l \hmu_l\cdot\vec\del\right)^2 \delta_{mn}
-\left(\gamma_m \hmu_m\cdot\vec\del\right)
 \left(\gamma_n \hmu_n\cdot\vec\del\right)\Bigr]T_n(\bfx). 
\end{equation}
We immediately identify the $S_m$'s are scalar fields in the continuum limit, 
and therefore impose 
\begin{equation}
 \sum_{m=1}^2 \Bigl(\gamma_m \hmu_m\cdot\vec\del\Bigr)^2 = l^2 \del^2, 
\label{Lorentz condition}
\end{equation}
for some constant $l$. 
This equation is easily solved by $\gamma_1=\gamma_2=l$ and 
\begin{equation}
 \hmu_1=\left(\begin{matrix}1 \\ 0 \end{matrix}\right), \qquad 
 \hmu_2=\left(\begin{matrix}0 \\ 1 \end{matrix}\right), 
 \label{mu 4SUSY}
\end{equation}
up to rotations and reflections. 
The kinetic terms (\ref{kinetic term 1}) then become 
\begin{equation}
-l^2S_m(\bfx)\del^2 S_m(\bfx)
 -l^2T_m(\bfx)\Bigl[\del^2\delta_{mn}
-\del_m\del_n\Bigr]T_n(\bfx). 
\end{equation}
The factor $l^2$ can be absorbed into the coupling constant 
by rescaling the fields appropriately, 
then they are the standard kinetic terms for scalar fields 
and a gauge field.

\begin{figure}[t]
  \begin{center}
    \includegraphics[scale=.7]{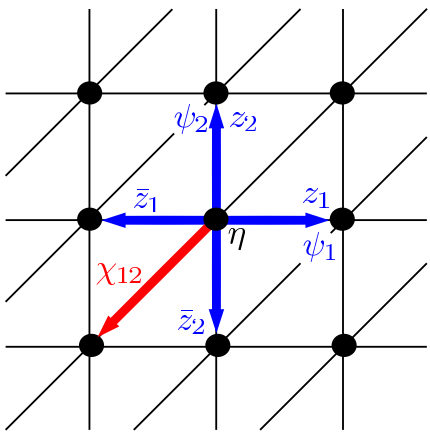}
  \end{center}
  \caption{The lattice structure and the field configuration of the
theory constructed from the mother theory with four supersymmetries.
 The lattice is essentially a square lattice and there are also diagonal
 links. $z_m$, $\bar{z}_m$, $\psi_m$ are on the usual link,  
 $\eta$ is on the site and $\chi_{12}$ is on the diagonal link.}
  \label{4SUSY_d=D=2}
\end{figure}

One can easily check that the obtained lattice action with 
the choice of (\ref{mu 4SUSY})
is identical to the one given in \cite{Cohen:2003xe}. Although different $a_m$
are allowed, they do not generate theories that are different from the one
of the natural choice $a_1=a_2=a$. 
Therefore, we conclude that the construction of that paper is the unique orbifolded 
lattice gauge theory starting from the mother theory (\ref{Smother0}). 
Different values of $a_m$ all give rise to the same naive continuum limit,
on account of eq. (\ref{continuum limit of difference operation}), and
only redefine what is meant by distance in the different directions.
In this formulation, $z_m$, $\bar{z}_m$, $\psi_m$ live on links, 
$\eta$ on sites and $\chi_{12}$ on diagonal links 
(see Fig. \ref{4SUSY_d=D=2}). Alternatively, one can think of $\chi_{12}$ as living
on corners.
As we have seen, in the naive continuum limit the real and imaginary 
components of $z_m$ become two scalar fields and a vector field,
respectively. 
Similarly, the fermion fields build up a two-dimensional Dirac
spinor in the continuum limit. 
As a result, it is expected that four supersymmetries are recovered and
the theory becomes two-dimensional $\cN=(2,2)$ supersymmetric 
gauge theory in the continuum limit.

\section{Target theories with eight supercharges}

A theory with eight supercharges is far reacher, and the classification of 
associated supersymmetric lattice action correspondingly more involved.
The mother theory can be obtained by dimensional reduction of six-dimensional
${\cal N}=1$ supersymmetric Yang-Mills theory, and in this case we take
the gauge group to be $U(kN^d)$ where $d$ can be either 2 or 3, the maximal
dimensionality of the lattice theory. In the notation of
\cite{Cohen:2003qw}, it can be written as 
\be
S_{\rm m} ~=~ \frac{1}{g^2}{\rm Tr}\left(-\frac{1}{4}[v_\alpha,v_\beta]^2 
+ \bar{\psi} \bar{\Sigma}_\alpha
[v_\alpha,\psi]\right) ~, \qquad (\alpha,\beta=0,\cdots,5)
\label{mather action for 8 SUSY}
\ee
where $\bar\Sigma_\alpha$ are defined through $SO(6)$ Dirac
matrices $\Gamma_\alpha$ as 
\begin{equation}
 \Gamma_\alpha = \left(\begin{matrix}0 & \Sigma_\alpha \\ 
 \bar\Sigma_\alpha & 0\end{matrix}\right). 
\end{equation}
$v_\alpha$ are $kN^d\times kN^d$ hermitian matrices and 
$\psi$ and $\bar{\psi}$ are independent complex four-component spinors. 
In the following, we use the representation,  
\begin{align}
 \bar\Sigma_0 &= 1_2\otimes 1_2,
 \quad \bar\Sigma_1 = -i \tau_3\otimes 1_2, 
 \quad \bar\Sigma_2 = i \tau_1\otimes\tau_1, \nn \\
 \bar\Sigma_3 &= -i \tau_1\otimes \tau_2,
 \quad \bar\Sigma_4 = -i \tau_1\otimes \tau_3, 
 \quad \bar\Sigma_5 = i \tau_2\otimes 1_2, 
\end{align}
together with
\begin{align}
 z_1 &\equiv v_0+iv_1, \quad z_2\equiv -i(v_2+iv_3), 
 \quad z_3 \equiv -i(v_4+iv_5), \nn \\
 \bar{z}_m &\equiv z_m^\dagger, \quad (m=1,2,3)
\end{align}
and 
\begin{equation}
 \psi \equiv \left(\begin{matrix}\eta\\ \xi_{23} \\ 
 \xi_{31} \\ \xi_{12}\end{matrix}\right), \qquad 
 \bar\psi^T \equiv \left(\begin{matrix}
 -\psi_1 \\ \chi_{123} \\ \psi_3 \\ -\psi_2
 \end{matrix}\right).  
\label{fermion configuration}
\end{equation}
Then the mother theory can be written as
\bea
S_{\rm m} &~=~& \frac{1}{g^2}\Bigl(
\frac{1}{4}|[z_m,z_n]|^2 
+\frac{1}{8}[z_m,z_m]^2 \cr
&& ~~~~~~
- \psi_m[\bar{z}_m,\eta]
+ \xi_{mn}[z_m,\psi_n]
+ \frac{1}{2}\chi_{lmn}[\bar{z}_l,\xi_{mn}]
\Bigr), \quad (l,m,n=1,2,3)
\label{mother action 2 for 8 SUSY}
\eea
where $\xi_{mn}$ and $\chi_{lmn}$ are completely antisymmetric 
with respect to the indices. 

The next step consists in identifying the maximal set of $U(1)$ symmetries.
As explained in ref. \cite{Cohen:2003qw}, the global symmetry of the mother
theory is $SO(6)\times SU(2)$ where $SO(6)$ is associated with the
Lorentz symmetry of the original six-dimensional theory, and $SU(2)$ is
a symmetry which acts only on the fermion fields. 
We then choose the maximal $U(1)$ symmetry as
\begin{equation} 
SO(6)\times SU(2)\supset SO(2)_{01}\times SO(2)_{23}\times 
SO(2)_{45}\times U(1), 
\end{equation}
where $SO(2)_{\alpha\beta}$ is the rotation group in the
$(\alpha,\beta)$-plane and $U(1)$ is the Cartan subgroup of $SU(2)$. 
The configuration of the $U(1)$ charges are summarized in
Table. \ref{charge table 8 SUSY} where $q_1\cdots q_4$ are the $U(1)$
charges corresponding to $SO(2)_{01}$, $SO(2)_{23}$, $SO(2)_{45}$ and
$U(1)$, respectively. 

\begin{table}[h]
\caption{The charge assignment of the maximal $U(1)$ symmetries for
 the component fields of the mother theory with eight supercharges} 
\begin{center}
\begin{tabular}{c|ccccccccccc}
 & $z_1$ & $z_2$ & $z_3$ & $\eta$ & $\xi_{23}$ & $\xi_{31}$ & $\xi_{12}$ 
 & $\chi_{123}$ & $\psi_1$ & $\psi_2$ & $\psi_3$  \\
\hline
$q_1$ & 1 & 0 & 0 & 1/2 & 1/2 & -1/2 & -1/2 & 1/2 & 1/2 & -1/2 & -1/2 \\
$q_2$ & 0 & 1 & 0 & 1/2 & -1/2 & 1/2 & -1/2 & 1/2 & -1/2 & 1/2 & -1/2 \\
$q_3$ & 0 & 0 & 1 & 1/2 & -1/2 & -1/2 & 1/2 & 1/2 & -1/2 & -1/2 & 1/2 \\
$q_4$ & 0 & 0 & 0 & 1/2 & 1/2 & 1/2 & 1/2 & -1/2 & -1/2 & -1/2 & -1/2 
\end{tabular}
\end{center}
\label{charge table 8 SUSY}
\end{table}

In order to make an orbifold projection, we need a set of $Z_N$
symmetries which are constructed by combining 
subgroups of the above four $U(1)$ symmetries. 
As in the previous section, our purpose is to construct 
all lattice theories which possess at least one scalar supersymmetry. 
We then need at least one fermion that is singlet under the $Z_N$
symmetries. We choose it to be $\eta$. Again the result is,
after relabelling, the same if
we choose another component of fermions as singlet. 
By this constraint, we can define at most three $U(1)$ symmetries 
for which all fields have integer charges as 
\begin{equation}
 r_m \equiv l_m^1 q_1 + l_m^2 q_2 + l_m^3 q_3 -(l_m^1+l_m^2+l_m^3) q_4, 
\qquad (m=1,2,3,\quad l_m^n\in \Z)
\end{equation}
and we define
\begin{equation}
 \fate_m \equiv \left(\begin{matrix}
                       l_1^m \\ l_2^m \\ l_3^m
                      \end{matrix}\right).\qquad (m=1,2,3)
\end{equation}
The integer charges for the component fields can be written 
in the form of a three-vector as 
\begin{equation}
 \bfr = q_1 \fate_1 + q_2 \fate_2 + q_3 \fate_3 - q_4
  (\fate_1+\fate_2+\fate_3), 
\end{equation}
which is summarized in Table \ref{integer charge table 8 SUSY}. 
Note that we do not assume that the $\fate_m$'s are linearly
independent. As expected, the dimensionality of the
lattice theory is determined by the number of linearly independent
components in $\{\fate_m\}$. 

\begin{table}[h]
\caption{The remaining three $U(1)$ charges}
\begin{center}
\begin{tabular}{c|ccccccccccc}
 & $z_1$ & $z_2$ & $z_3$ & $\eta$ & $\xi_{23}$ & $\xi_{31}$ & $\xi_{12}$ 
 & $\chi_{123}$ & $\psi_1$ & $\psi_2$ & $\psi_3$  \\
\hline
 $\bfr$ & $\fate_1$ & $\fate_2$ & $\fate_3$ 
& 0 & $-\fate_2-\fate_3$ & -$\fate_3-\fate_1$ & $-\fate_1-\fate_2$ 
& $\fate_1+\fate_2+\fate_3$ & $\fate_1$ & $\fate_2$ & $\fate_3$ \\
\end{tabular}
\end{center}
\label{integer charge table 8 SUSY}
\end{table}

We can then carry out the orbifold
projection by a $Z_N^d$ subgroup of the remaining three $U(1)$
symmetries. 
As in the previous section, it can be achieved by expanding 
all fields by fields living on an abstract lattice labelled by 
vectors $\fatk$:
\bea
z_m &~=~& \sum_{\fatk}z_m(\fatk)\otimes E_{\fatk,\fatk+\fate_m} \cr
\bar{z}_m   &~=~& \sum_{\fatk}\bar{z}_m(\fatk)\otimes E_{\fatk+\fate_m,\fatk} \cr
\eta  &~=~& \sum_{\fatk}\eta(\fatk)\otimes E_{\fatk,\fatk} \cr
\psi_m &~=~& \sum_{\fatk}\psi_m(\fatk)\otimes E_{\fatk,\fatk+\fate_m} \cr
\xi_{mn} &~=~&
\sum_{\fatk}\xi_{mn}(\fatk)\otimes E_{\fatk+\fate_m+\fate_n,\fatk} \cr 
\chi_{123} &~=~&
\sum_{\fatk}\chi_{123}(\fatk)\otimes
E_{\fatk,\fatk+\fate_1+\fate_2+\fate_3}, 
\label{expansion}
\eea
where $z_m(\fatk)$, $\bar{z}_m(\fatk)$, $\cdots$ are the fields on the
abstract lattice and $E_{\bfk,{\mathbf l}}$ is defined in Appendix A. 
Substituting this expansion into the action 
(\ref{mother action 2 for 8 SUSY}), we obtain the action for the 
orbifolded theory; 
\bea
S_{\rm orb} &=& \frac{1}{g^2}{\rm Tr}\sum_{\mathbf k}\Biggl( 
\frac{1}{4}\Bigl| z_m({\mathbf k})
z_n(\fatk+\fate_m) - z_n(\fatk)z_m(\fatk+\fate_n)\Bigr|^2  \cr
&& + \frac{1}{8}\Bigl(z_m(\fatk)\bar{z}_m(\fatk)-\bar{z}_m(\fatk-\fate_m)z_m(\fatk
-\fate_m)\Bigr)^2 \cr
&& - \psi_m(\fatk)\Bigl(\bar{z}_m(\fatk)\eta(\fatk)-\eta(\fatk+\fate_m)
\bar{z}_m(\fatk)\Bigr)\cr 
&& +\frac{1}{2} \xi_{mn}(\fatk)\Bigl(z_m(\fatk)\psi_n(\fatk+\fate_m)
-\psi_n(\fatk)z_m(\fatk+\fate_n)-(m\!\leftrightarrow\!n)\Bigr) \cr
&& -\frac{1}{2}\chi_{lmn}(\bfk)\Bigl(
\bar{z}_l(\bfk+\bfe_m+\bfe_n)\xi_{mn}(\bfk)
-\xi_{mn}(\bfk+\bfe_l)\bar{z}_l(\bfk)
\Bigr)\Biggr)~.  \cr
&&
\label{orbifold action 8 SUSY}
\eea
This orbifolded theory has $U(k)$ ``gauge symmetry''; 
\begin{alignat}{2}
 z_m(\bfk) &\to V(\bfk)^\dagger z_m(\bfk)V(\bfk+\bfe_m), &
 \bar{z}_m(\bfk) &\to V(\bfk+\bfe_m)^\dagger \bar{z}_m(\bfk)V(\bfk), \nn
 \\
 \psi_m(\bfk) &\to V(\bfk)^\dagger \psi_m(\bfk)V(\bfk+\bfe_m), &
 \bar{\psi}_m(\bfk) &\to V(\bfk+\bfe_m)^\dagger \bar{\psi}_m(\bfk)V(\bfk), \nn
 \\
 \eta(\bfk) &\to V(\bfk)^\dagger \eta(\bfk) V(\bfk), &
 \xi_{mn}(\bfk) &\to V(\bfk+\bfe_m+\bfe_n)^\dagger \xi_{mn}(\bfk) V(\bfk),
 \nn \\
 \chi_{123}(\bfk) &\to V(\bfk)^\dagger \chi_{123}(\bfk)
 V(\bfk+\bfe_1+\bfe_2+\bfe_3),
\end{alignat}
with $V(\bfk)\in U(k)$. 
Based on the orbifolded action (\ref{orbifold action 8 SUSY}), 
we will construct three-dimensional and two-dimensional lattice theories
in turn. 
As mentioned above, the dimensionality of the lattice equals  
the number of linearly independent vectors in $\{\bfe_m\}$.

\subsection{Three-dimensional Lattice ($d=3$)}

In this subsection we assume that all $\bfe_m$ are linearly independent so that 
$\{\bfe_m\}$ forms a basis of a three-dimensional lattice. 
As in the previous section, we shift $z_m$ (and $\bar{z}_m$) 
in order to generate kinetic terms. 
Since we want to construct a three-dimensional theory, we must shift all three $z_m$ 
so that $(N_{\rm shift}=3)$:
\begin{equation}
 z_m(\bfk) \to \frac{1}{a_m} + z_m(\bfk). \qquad (m=1,2,3)
\end{equation}
We can again assume $a_m\in\R_+$ without loss of generality. 
Then the lattice action becomes 
\bea
S_{\rm lat}^{d=3,N=3} &=& \frac{1}{g^2}{\rm Tr}\sum_{\mathbf k}\Biggl( 
\frac{1}{4}\Bigl| 
\nabla_m^+ z_n(\bfk) - \nabla_n^+ z_m(\bfk)
+z_m({\mathbf k})z_n(\fatk+\bfe_m) 
- z_n(\fatk)z_m(\fatk+\bfe_n)\Bigr|^2  \cr
&& + \frac{1}{8}\Bigl(
\nabla_m^+\left(z_m(\bfk)+\bar{z}_m(\bfk)\right)
+z_m(\fatk+\bfe_m)\bar{z}_m(\fatk+\bfe_m)
-\bar{z}_m(\fatk)z_m(\fatk)\Bigr)^2 \cr
&& - \psi_m(\fatk)\Bigl(
\nabla_m^+ \eta(\bfk)
-\bar{z}_m(\fatk)\eta(\fatk)
+\eta(\fatk+\bfe_m)\bar{z}_m(\fatk)\Bigr)\cr 
&& + \frac{1}{2}\xi_{mn}(\fatk)\Bigl(
\nabla_m^+ \psi_n(\bfk)
+z_m(\fatk)\psi_n(\fatk+\bfe_m)
-\psi_n(\fatk)z_m(\fatk+\bfe_n)
-(m\!\leftrightarrow\!n)\Bigr) \cr
&& -\frac{1}{2}\chi_{lmn}(\bfk)\Bigl(
\nabla_l^+ \xi_{mn}(\bfk)
-\bar{z}_l(\bfk+\bfe_m+\bfe_n)\xi_{mn}(\bfk)
+\xi_{mn}(\bfk+\bfe_l)\bar{z}_l(\bfk) 
\Bigr)\Biggr)~. \cr
&&
\label{d=3 D=3 lattice action}
\eea
where the difference operator $\nabla_m^+$ is defined by 
(\ref{difference operation}).
As for the case of the four supercharges, the lattice theory is labelled 
by the values of $a_m$ and the linear relation among $\bfe_m$. 
Therefore, since the $\bfe_m$ by construction are linearly independent
here, the three-dimensional lattice theory depends only on the parameters 
$a_m$.

\begin{figure}[t]
  \begin{center}
    \includegraphics[scale=1]{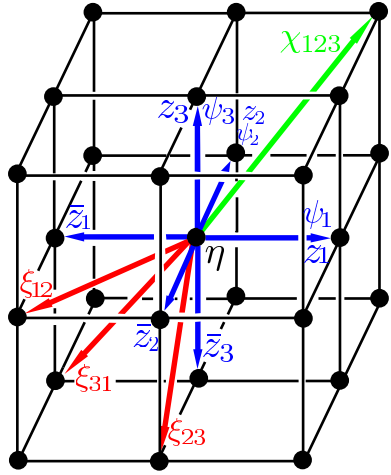}
  \end{center}
  \caption{The lattice structure and the field configuration of the
 three-dimensional lattice formulation.}
  \label{8SUSY_d=D=3}
\end{figure}

Next we consider the continuum limit. 
Again, we introduce lattice spacings through the linear map, 
\begin{equation}
 f: \bfe_m \mapsto \gamma_m a_m \hmu_m, \qquad (m=1,2,3)
\end{equation}
and define the continuum limit by $a_m\to 0$. 
The difference operator $\nabla_m^+$ 
becomes a derivative operator $\gamma_m \hmu_m\cdot\vec\del$ in that limit. 
As in the previous section 
the condition of Lorentz invariance in the continuum limit determines 
the set $\{\hmu_m\}$ uniquely as 
\begin{equation}
 \hmu_1=\left(\begin{matrix}
               1 \\ 0 \\ 0
              \end{matrix}\right), \qquad
 \hmu_2=\left(\begin{matrix}
               0 \\ 1 \\ 0
              \end{matrix}\right), \qquad
 \hmu_3=\left(\begin{matrix}
               0 \\ 0 \\ 1
              \end{matrix}\right), \qquad
\label{Kaplan's mu for d=D=3}
\end{equation}
up to rotations and reflections, 
and $\gamma_1=\gamma_2=\gamma_3$.  
In this way we obtain the action that was  
constructed in \cite{Cohen:2003qw} 
after setting all $a_m$ equal.
The lattice is cubic and 
the theory possesses one scalar supersymmetry from the
orbifold construction. Variables
$z_m$, $\bar{z}_m$, $\psi_m$ live on the usual links, 
$\eta$ on sites, $\xi_{mn}$ on square diagonal links, 
and $\chi_{123}$ lives on body diagonal links (see Fig.\ref{8SUSY_d=D=3}). 
In the continuum limit, the real and imaginary components of $z_m$
become three scalar fields and a vector field, respectively. 
The eight fermion fields form two three-dimensional 
Dirac spinors. As a result, eight supersymmetries are recovered 
in the naive continuum limit which turns out to yield
three-dimensional supersymmetric gauge theory with 8 supersymmetries 
\cite{Cohen:2003qw}. 
We conclude that (\ref{d=3 D=3 lattice action}) is the unique
three-dimensional lattice action constructed by orbifolding from the mother theory
given in (\ref{mather action for 8 SUSY}).

\subsection{Two-dimensional Lattice ($d=2$)}

In this subsection, we classify the two-dimensional lattices 
that can be constructed 
from the mother theory (\ref{mather action for 8 SUSY}).
Two dimensional theories appear when there are only two linearly
independent vectors in $\{\bfe_m\}$, which we choose to be 
$\bfe_1$ and $\bfe_2$. The vector $\bfe_3$ can thus be expressed as 
a linear combination of these: 
\begin{equation}
 \bfe_3 = p \bfe_1 + q \bfe_2. \qquad (p,q\in\Q)
 \label{linear relation}
\end{equation}
The fact that $p$ and $q$ must be rational numbers follows from the
quantization of the $U(1)$ charges and the definition of the ${\bfe_m}$'s.
For the purpose of the future discussion, we assume that 
$p$ and $q$ satisfy 
\begin{equation}
 pq \ge 0.
\label{condition for sign}
\end{equation}
Otherwise, we can always swap the roles of $\bfe_m$ so that 
they satisfy (\ref{condition for sign}). 
The new point in the present case is 
that we can construct a lattice theory 
by shifting either two ($N_{\rm shift}=2$) or three $(N_{\rm shift}=3)$ of $z_m$, 
and this gives rise to different lattice theories. 
As we shall see below, one can in fact construct seven distinct 
groups of lattice theories that all correspond to 
two-dimensional $\cN=(4,4)$ gauge theory in the naive continuum limit. 
Five of these lattice theories 
have two conserved supercharges while the remaining two have only one.

\subsubsection{$N_{\rm shift}=2$}
\label{Ns=2}

We first we consider the case where one shifts only $z_1$ and $z_2$; 
\begin{align}
 z_m(\bfk) &\to \frac{1}{a_m} + z_m(\bfk), \quad (m=1,2) \nn \\
 z_3(\bfk) &\to z_3(\bfk), 
\end{align}
with $a_m\in \R_+$ $(m=1,2)$. 
Then we obtain the lattice action, 
\bea
S_{\rm lat}^{d=2,N=2} &=& 
\frac{1}{g^2}{\rm Tr}\sum_{\mathbf k}\Biggl( 
\frac{1}{4}\Bigl| 
\nabla_m^+ z_n(\bfk) - \nabla_n^+ z_m(\bfk)
+z_m({\mathbf k})z_n(\fatk+\bfe_m) 
- z_n(\fatk)z_m(\fatk+\bfe_n)\Bigr|^2  \cr
&& + \frac{1}{8}\Bigl(
\nabla_m^+\left(z_m(\bfk)+\bar{z}_m(\bfk)\right)
+z_m(\fatk+\bfe_m)\bar{z}_m(\fatk+\bfe_m)-\bar{z}_m(\fatk)z_m(\fatk)
\Bigr)^2 \cr
&& +\frac{1}{2}\Bigl|
\nabla_m^+ z_3(\bfk) + z_m(\bfk)z_3(\bfk+\bfe_m)-z_3(\bfk)z_m(\bfk+\bfe_3)
\Bigr|^2 \cr
&& +\frac{1}{8}\Bigl(
z_3(\fatk)\bar{z}_3(\fatk)-\bar{z}_3(\fatk-\bfe_3)z_3(\fatk-\bfe_3)
\Bigr)^2 \cr
&& - \psi_m(\fatk)\Bigl(
\nabla_m^+ \eta(\bfk)
-\bar{z}_m(\fatk)\eta(\fatk)+\eta(\fatk+\bfe_m)
\bar{z}_m(\fatk)\Bigr)\cr 
&& - \psi_3(\fatk)\Bigl(
\bar{z}_3(\fatk)\eta(\fatk)-\eta(\fatk+\bfe_3)
\bar{z}_m(\fatk)\Bigr)\cr 
&& +\frac{1}{2} \xi_{mn}(\fatk)\Bigl(
\nabla_m^+\psi_n(\bfk)
+z_m(\fatk)\psi_1(\fatk+\bfe_n) -\psi_n(\fatk)z_2(\fatk+\bfe_m) 
-(m\!\leftrightarrow\!n)\Bigr) \cr
&& + \frac{1}{2}\xi_{m3}(\fatk)\Bigl(
\nabla_m^+ \psi_3(\bfk)
+z_m(\fatk)\psi_3(\fatk+\bfe_m)-\psi_3(\fatk)z_m(\fatk+ \bfe_3) \cr
&&\hspace{3.6cm}
-z_3(\fatk)\psi_m(\fatk+\bfe_3)+\psi_m(\fatk)z_3(\fatk+\bfe_m) 
\Bigr) \cr
&& -\chi_{123}(\bfk)\biggl(
\e_{mn}\Bigl(
\nabla_m^+ \xi_{n3}
-\bar{z}_m(\bfk+\bfe_n+\bfe_3)\xi_{n3}(\bfk)
+\xi_{n3}(\bfk+\bfe_m)\bar{z}_m(\bfk)
\Bigr) \cr
&&\hspace{3cm}-\bar{z}_3(\bfk+a\hmu_1+a \hmu_2)\xi_{12}(\bfk)
+\xi_{12}(\bfk+\bfe_3)\bar{z}_3(\bfk)
\biggr)\Biggr)~, \cr
&&
\label{d=2 D=2 lattice action}
\eea
with an implicit summation over $m,n=1,2$. 
Again, the lattice theory is labelled by the values of the $a_m$ 
and the linear relation among the vectors $\bfe_m$, $i.e.$, by the values of 
$p$ and $q$ in (\ref{linear relation}). We thus
obtain infinitely many lattice formulations in this case. 
As we will see below, they can be classified by the number of 
preserved supersymmetries on the lattice.

We again introduce lattice spacings through a linear mapping, 
\begin{equation}
 f: \bfe_m \mapsto \gamma_m a_m \hmu_m,\quad  |\hmu_m|=1,
  \quad \gamma_m\in\R_+  
\qquad (m=1,2)
\end{equation}
and define the continuum limit by $a_m\to 0$ ($m=1,2$) as before. 
Since $f$ is linear, $\bfe_3$ is mapped to 
$p \gamma_1 a_1 \hmu_1 + q \gamma_2 a_2 \hmu_2$. 
In this case, repeating the proof at the end of Section \ref{Target
theories with four supercharges},  
we can show that the continuum theory can be Lorentz 
invariant if and only if 
\begin{align}
  \hmu_1=\left(\begin{matrix}
               1 \\ 0
              \end{matrix}\right), \qquad 
  \hmu_2=\left(\begin{matrix}
               0 \\ 1
              \end{matrix}\right),
\end{align}
up to rotations and reflections, and $\gamma_1=\gamma_2$. 
This is a square lattice. 

Although there are infinitely many lattice formulations 
labelled by $(p,q)$, we can classify them by 
the number of remaining supersymmetries. 
In fact, 
this number can be enhanced by tuning $\bfe_3$ properly. There are three cases.

\begin{figure}[t]
  \begin{center}
   \parbox{.4\linewidth}{\begin{center}
    \includegraphics[scale=1]{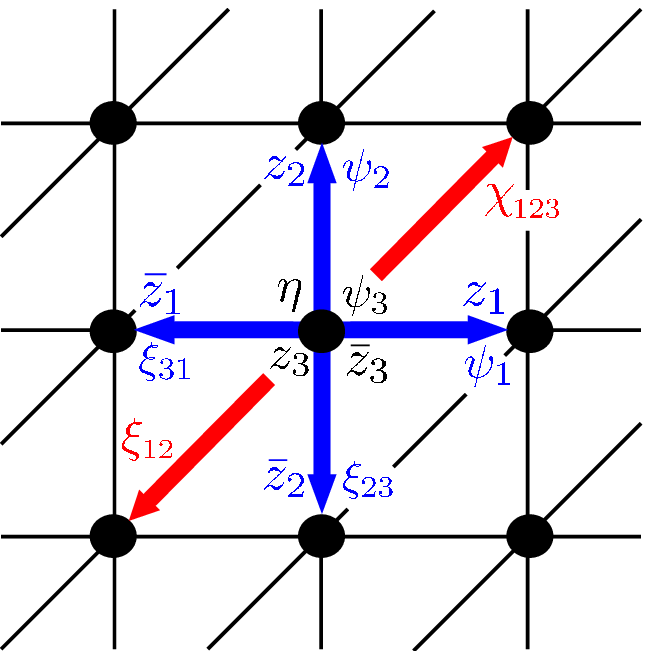}\\
    (a) $\bfe_3=0$\end{center}}
\hspace{1cm}
    \parbox{.4\linewidth}{\begin{center}
    \includegraphics[scale=1]{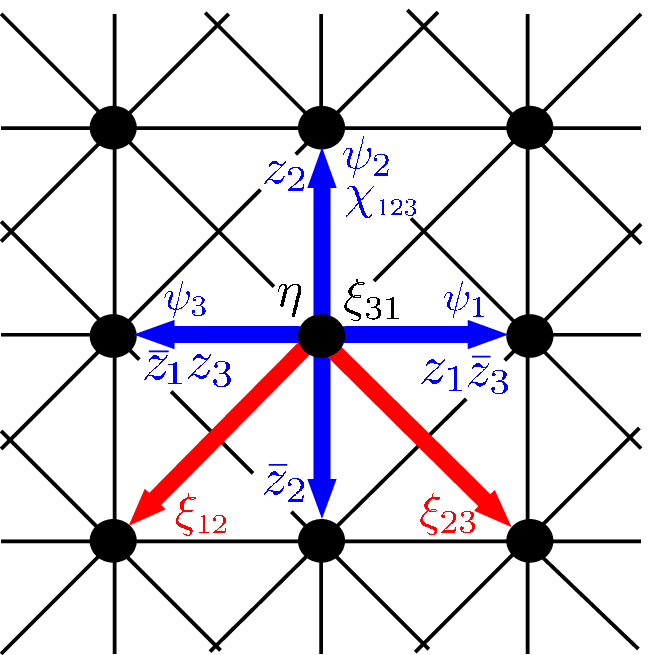}\\
    (b) $\bfe_3=-\bfe_1$\end{center}}
  \end{center}
  \caption{The lattice structure and the field configuration of the
 two-dimensional lattice formulation constructed
 by shifting $z_1$ and $z_2$. 
 We have drawn the case of (a) $\bfe_3=0$ and (b) $\bfe_3=-\bfe_1$, 
 where there are two remaining supercharges.  
 Correspondingly, there are two fermions on sites for the both cases.}
  \label{8SUSY_d=D=2 Q=2}
\end{figure}

\vspace{3mm}
\noindent
[1-1] $\bfe_3=0$ 

This is the two-dimensional theory constructed 
in \cite{Cohen:2003qw}, where it was already 
shown that it possesses two scalar supersymmetries.  
In fact, looking at Table \ref{integer charge table 8 SUSY}, 
We see that $\psi_3$ also becomes a scalar fermion 
by this choice of $\bfe_3$. 
In this formulation, $z_3$, $\bar{z}_3$, $\eta$ and $\psi_3$ live on 
sites, $z_m$, $\bar{z}_m$ and $\psi_m$ $(m=1,2)$ live on the usual links, 
and $\xi_{12}$ and $\chi_{123}$ are on diagonal links (see (a) 
in Fig. \ref{8SUSY_d=D=2 Q=2}). 
In the continuum limit, the real components of $z_m$ 
($m=1,2$) and $z_3$ become four real scalar fields, 
the imaginary components of $z_m$ $(m=1,2)$ becomes a vector field, 
and the fermion fields combine into two two-dimensional Dirac spinors. 
As a result, 
as discussed in \cite{Cohen:2003qw}, the continuum theory is 
expected to be two-dimensional 
$\cN=(4,4)$ supersymmetric gauge theory.

\vspace{3mm}
\noindent
[1-2] $\bfe_3=-\bfe_1$  (${\rm or}\ -\bfe_2$) 

This gives a new lattice formulation of two-dimensional 
$\cN=(4,4)$ supersymmetric gauge theory.
As for the case of $\bfe_3=0$, there is an ``accidental''
enhancement of supersymmetries and there are again two conserved
supercharges on the lattice.
In fact, looking at Table \ref{integer charge table 8 SUSY}, 
we see that $\xi_{31}$ (or $\xi_{23}$) becomes a singlet under the
$U(1)$ transformations. We thus expect on
general grounds that there will be two preserved supersymmetries
in this case. This can indeed be checked explicitly, both for
this case and for the subsequent cases discussed below. We have
summarized the proof of this in Appendix B.
In this case, $\eta$ and $\xi_{31}$ live on sites, while
$z_m$, $\bar{z}_m$, $\psi_m$ ($m=1,2,3$) and $\chi_{123}$ 
live on the usual links,  
and $\xi_{12}$ and $\xi_{23}$ sit on diagonal links 
but in the direction opposite of the case corresponding to $\bfe_3=0$ 
(see (b) of Fig. \ref{8SUSY_d=D=2 Q=2}).
The role of the fields in the continuum limit is completely the same as
in the case of $\bfe_3=0$, and the continuum theory is again
expected to be two-dimensional $\cN=(4,4)$ supersymmetric gauge theory. 

\vspace{3mm}
\noindent
[1-3] $\bfe_1+\bfe_2+\bfe_3=0$

\begin{figure}[t]
  \begin{flushleft}
   \parbox{.4\linewidth}{\begin{center}
    \includegraphics[scale=.7]{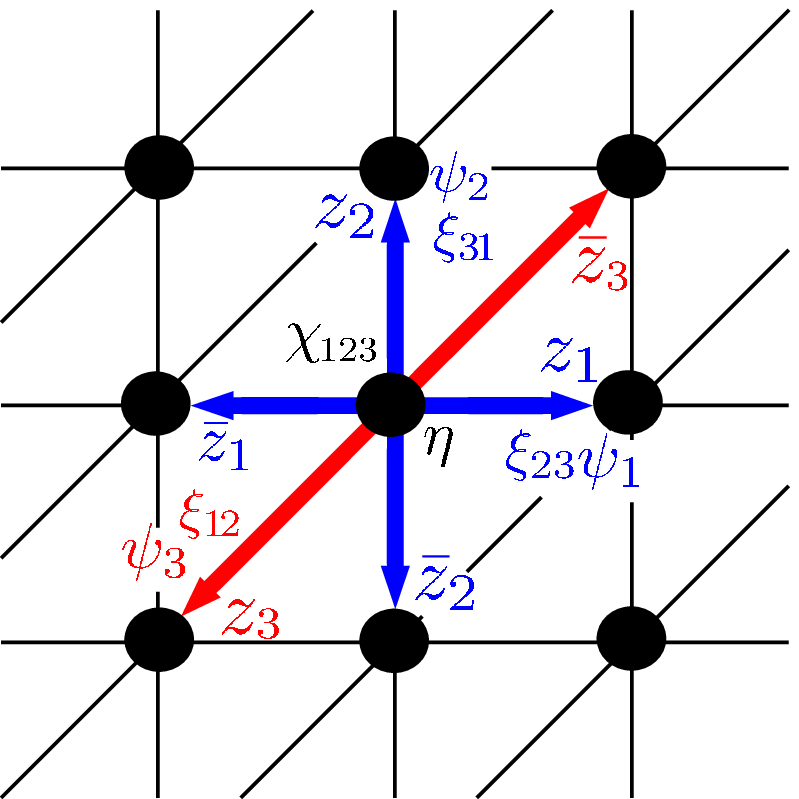}\\
    (c) $\bfe_1+\bfe_2+\bfe_3=0$\end{center}}
\hspace{.3cm}
    \parbox{.4\linewidth}{\begin{center}
   \includegraphics[scale=1.2]{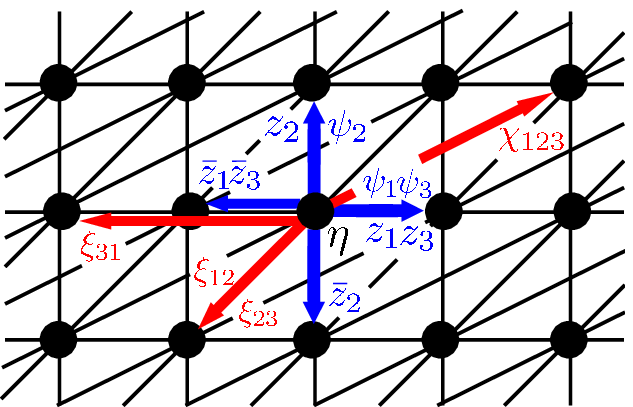}\\
    (d) $\bfe_3=\bfe_1$\end{center}}
  \end{flushleft}
  \caption{The lattice structure and the field configuration of 
 two-dimensional lattice formulation constructed by shifting $z_1$ and
 $z_2$. We have drawn the cases of 
 (c) $\bfe_1+\bfe_2+\bfe_3=0$ and (d) $\bfe_3=\bfe_1$. 
 There are two remaining supercharges in (c), while there is 
 one supercharge in (d). 
 Comparing to the three cases (a), (b) and (c), both of 
 the lattice structure and the field configuration are less symmetric
in the case of (d).}
  \label{8SUSY_d=D=2 Q=1}
\end{figure}

This also gives a new lattice formulation of 
two-dimensional $\cN=(4,4)$ supersymmetric gauge theory.
In this formulation, $\eta$ and $\chi_{123}$ live on sites, and there thus
two remaining supersymmetries as in [1-1] and [1-2] above. 
The fields $z_m$, $\psi_m$, $\xi_{23}$ and $\xi_{31}$ ($m=1,2$) 
live on links, 
and $z_3$, $\psi_3$ and $\xi_{12}$ live on diagonal links 
(see (c) of Fig. \ref{8SUSY_d=D=2 Q=1}).
The role of the fields in the naive continuum limit is again the same 
as in the cases of [1-1] and [1-2], 
and the continuum theory is expected 
 to be two-dimensional $\cN=(4,4)$ supersymmetric gauge theory.

\vspace{3mm}
\noindent
[1-4] $\bfe_3 \notin \{0,\, -\bfe_1,\,-\bfe_2,\, -\bfe_1-\bfe_2\} $

This is again a new lattice formulation of two-dimensional 
$\cN=(4,4)$ supersymmetric gauge theory, 
but there is now only one supersymmetry preserved on the lattice 
at finite lattice spacing. 
In this case, the structure of the lattice is less symmetric than the
three cases above and there are several kinds of diagonal links in
general. 
We draw the case of $\bfe_3=\bfe_1$ as an example in
(d) of Fig. \ref{8SUSY_d=D=2 Q=1}. 
Although there is less lattice (space-time) symmetry, 
the roles of the fields in the continuum limit 
are the same, and the naive continuum theory is again 
expected to be two-dimensional
$\cN=(4,4)$ supersymmetric gauge theory.

\subsubsection{$N_{\rm shift}=3$}
\label{Ns=3}

Next, let us consider the case where we shift all $z_m$: 
\begin{align}
 z_m(\bfk) \to \frac{1}{a_m} + z_m(\bfk). \qquad (m=1,2,3)
 \label{shift of zm}
\end{align}
Then we obtain the action, 
\bea
S_{\rm lat}^{d=2,N=3} &=& 
\frac{1}{g^2}{\rm Tr}\sum_{\mathbf k}\Biggl( 
\frac{1}{4}\Bigl| 
\nabla_m^+ z_n(\bfk) - \nabla_n^+ z_m(\bfk)
+z_m({\mathbf k})z_n(\fatk+\bfe_m) 
- z_n(\fatk)z_m(\fatk+\bfe_n)\Bigr|^2  \cr
&& + \frac{1}{8}\Bigl(
\nabla_m^+\left(z_m(\bfk)+\bar{z}_m(\bfk)\right)
+z_m(\fatk+\bfe_m)\bar{z}_m(\fatk+\bfe_m)
-\bar{z}_m(\fatk)z_m(\fatk)\Bigr)^2 \cr
&& - \psi_m(\fatk)\Bigl(
\nabla_m^+ \eta(\bfk)
-\bar{z}_m(\fatk)\eta(\fatk)
+\eta(\fatk+\bfe_m)\bar{z}_m(\fatk)\Bigr)\cr 
&& + \frac{1}{2}\xi_{mn}(\fatk)\Bigl(
\nabla_m^+ \psi_n(\bfk)
+z_m(\fatk)\psi_n(\fatk+\bfe_m)
-\psi_n(\fatk)z_m(\fatk+\bfe_n)
-(m\!\leftrightarrow\!n)\Bigr) \cr
&& -\frac{1}{2}\chi_{lmn}(\bfk)\Bigl(
\nabla_l^+ \xi_{mn}(\bfk)
-\bar{z}_l(\bfk+\bfe_m+\bfe_n)\xi_{mn}(\bfk)
+\xi_{mn}(\bfk+\bfe_l)\bar{z}_l(\bfk) 
\Bigr)\Biggr)~, \cr
&&
\label{d=2 D=3 lattice action}
\eea
with implicit summation over $l,m,n=1,2,3$. 
This action is formally of the same form as the
three-dimensional lattice theory (\ref{d=3 D=3 lattice action}), 
but the interpretation is completely different because now the 
three vectors $\bfe_m$ span a two-dimensional space-time. 
As for the above cases,
the lattice formulation is labelled by the values of 
$a_m$'s ($m=1,2,3$) and the values of $p$ and $q$. 

Again, we introduce lattice spacings through the mapping, 
\begin{equation}
 f:\bfe_m \mapsto \gamma_m a_m \hmu_m, \quad |\hmu_m|=1, \qquad
  (m=1,2,3)
\label{linear mapping for N=3}
\end{equation}
and the continuum limit is defined by $a_m\to 0$ ($m=1,2,3$). 
The condition for Lorentz invariance is the same as before,
\begin{equation}
 \sum_{m=1}^3 \Bigl(\gamma_m \hmu_m\cdot\vec\del\Bigr)^2 = l^2 \del^2, 
\label{Lorentz condition d=2 D=3}
\end{equation}
for some constant $l$. 
In this case, the condition of Lorentz invariance 
in the naive continuum limit does not determine $\hmu_m$ uniquely, 
but determines only the relation between $\hmu_m$ and $\gamma_m$. 
To see this, we write $\hmu_m$ as 
\begin{equation}
 \hmu_1=\left(\begin{matrix}1 \\ 0 \end{matrix}\right), \quad
 \hmu_2=\left(\begin{matrix}
   \cos \theta_2 \\ \sin \theta_2 \end{matrix}\right), \quad
 \hmu_3=\left(\begin{matrix}
   \cos \theta_3 \\ \sin \theta_3 \end{matrix}\right). \quad
 (0\le\theta_2\le\theta_3<2\pi)
\end{equation}
When $\sin\theta_2$, $\sin\theta_3$ and $\sin(\theta_3-\theta_2)$ are
all non-zero, the solutions for $\gamma_m$ are 
\begin{align}
 \gamma_1^2 &= \frac{l^2 \cos \theta_{32}}{\sin\theta_2\sin\theta_3}. 
 \quad
 \gamma_2^2 = \frac{-l^2}{\sin \theta_{32}}
              \frac{\cos\theta_3}{\sin\theta_2},
 \quad 
 \gamma_3^2 = \frac{l^2}{\sin \theta_{32}}
              \frac{\cos\theta_2}{\sin\theta_3},
\label{values of gamma}
\end{align}
where $\theta_{32}\equiv \theta_3-\theta_2$.
Since the $\gamma_m^2$ must be positive, 
the ranges of $\theta_1$ and $\theta_2$ are restricted to
\begin{align}
\label{one SUSY case}
  \frac{\pi}{2}<\theta_2 < \pi&,\quad 
  \theta_2+\frac{\pi}{2}<\theta_3<\frac{3\pi}{2}, \\
 &{\rm or} \nn \\
  \pi<\theta_2 < \frac{3\pi}{2}&, \quad
  \frac{3\pi}{2}<\theta_3<\theta_2+\frac{\pi}{2}, 
\label{two SUSY case}
\end{align}
up to rotations and flips of $\hmu_m$.
Here, we have required 
that $\hmu_3$ satisfy the same condition with (\ref{condition for
sign}), that is, $\hmu_3 = \alpha \hmu_1 + \beta \hmu_2$ with 
$\alpha\beta\ge 0$. 
At the boundary of these regions, we cannot use the formula 
(\ref{values of gamma}) and special care is needed. 
In this case, we obtain 
\begin{align}
 \gamma_1^2 &=l^2,\  \gamma_2^2+\gamma_3^2=l^2, \qquad
 (\theta_2,\theta_3)=(\frac{\pi}{2},\frac{\pi}{2}),
                     (\frac{\pi}{2},\frac{3\pi}{2}),
                     (\frac{3\pi}{2},\frac{3\pi}{2}) \nn \\
\label{boundary cases}
  \gamma_2^2 &=l^2, \ \gamma_1^2 +\gamma_3^2 =l^2,
\qquad (\theta_2,\theta_3)=(\frac{\pi}{2},\pi) \\
  \gamma_3^2 &=l^2, \ \gamma_1^2 +\gamma_2^2 =l^2.
\qquad (\theta_2,\theta_3)=({\pi},\frac{3\pi}{2}) \nn 
\end{align}

The situation is different from the previous cases in that 
there are now restrictions on the values of $a_m$. 
This can be seen as follows. 
By definition, $f$ maps $\bfe_m$ as 
\begin{align}
 f: 
\begin{cases}
   \bfe_1 &\mapsto \gamma_1 a_1\hmu_1 
 = \left(\gamma_1 a_1, 0\right) , 
 \\
 \bfe_2 &\mapsto \gamma_2 a_2\hmu_2 
= \left(\gamma_2 a_2 \cos \theta_2, \gamma_2 a_2 \sin \theta_2\right),
 \\
\bfe_3 &\mapsto \gamma_3 a_3\hmu_3 
= \left(\gamma_3 a_3 \cos \theta_3, \gamma_3 a_3 \sin \theta_3\right). 
\end{cases}
\label{maps1}
\end{align} 
On the other hand, since $f$ is a linear mapping, 
the combinations $a_m\hmu_m$ must also satisfy 
\begin{equation}
 \gamma_3 a_3\hmu_3 = p \gamma_1 a_1 \hmu_1 + q \gamma_2 a_2 \hmu_2. 
 \label{linear condition after mapping}
\end{equation}
Eqs. (\ref{maps1}) and (\ref{linear condition after mapping}) suggest
that $a_2$ and $a_3$ can be solved in terms of $a_1$: 
\begin{equation}
 a_2 = \frac{p}{q} \sqrt{-\frac{\tan\theta_3}{\tan(\theta_3-\theta_2)}}
       a_1, \qquad 
 a_3 = \pm p \sqrt{\frac{\tan \theta_2}{\tan(\theta_3-\theta_2)}}
       a_1, 
 \label{condition for am}
\end{equation}
where we take plus/minus sign in the case of 
(\ref{one SUSY case})/(\ref{two SUSY case}), respectively.
Since $a_m\in\R_+$, we see that there is a restriction 
on the regions of $(\theta_2, \theta_3)$ corresponding to the signature 
of $p$ and $q$;
if $p>0$ and $q>0$, we must use (\ref{one SUSY case}), and 
if $p<0$ and $q<0$, we must use (\ref{two SUSY case}) 
(recall eq. (\ref{condition for sign})). 
As we shall see, $\theta_2$ and $\theta_3$ take 
the values of (\ref{boundary cases}) in the cases of $p=0$ or $q=0$. 

Conversely, if we impose Lorentz invariance in the naive
continuum limit, the linear mapping $f$ is completely determined by 
given values of $(p,q)$ and $\{a_m\}$ up to the overall factor $l$. 
In fact, (\ref{condition for am}) can be inverted to give 
\begin{equation}
 \tan\theta_2 = \frac{a_3}{q a_2}
\sqrt{1+\frac{q^2a_2^2+a_3^2}{p^2a_1^2}}, \qquad 
\tan\theta_3 = - \frac{qa_2}{a_3}
\sqrt{1+\frac{q^2a_2^2+a_3^2}{p^2a_1^2}}, 
\label{values of theta}
\end{equation}
which determines $\theta_2$ and $\theta_3$ uniquely. 
The values of $\gamma_m$ are also determined 
through the relation (\ref{values of gamma}) as 
\begin{align}
 \gamma_1^2 &= \frac{q^2a_2^2 + a_3^2}{p^2a_1^2+q^2a_2^2+a_3^2} l^2, \nn
 \\
 \label{gamma wrt pqa}
 \gamma_2^2 &= \frac{p^2q^2a_1^2a_2^2 + a_3^2(p^2a_1^2+q^2a_2^2+a_3^2)}
  {(q^2a_2^2+a_3^2)(p^2a_1^2+q^2a_2^2+a_3^2)} l^2 , \\
 \gamma_3^2 &= \frac{p^2a_1^2a_3^2 + q^2a_2^2(p^2a_1^2+q^2a_2^2+a_3^2)}
 {(q^2a_2^2+a_3^2)(p^2a_1^2+q^2a_2^2+a_3^2)} l^2. \nn 
\end{align}
In particular, the formulae (\ref{values of theta}) and 
(\ref{gamma wrt pqa}) can be applied to the cases of 
$p=0$ or $q=0$, 
which give (\ref{boundary cases}) as announced. 
We exclude the case of $p=q=0$ here since it leads to a vanishing
$\gamma_3$.  
We will discuss this point in the next subsection. 

In summary, we can conclude that the lattice formulation 
for $N_{\rm shift}=3$ is labelled by 
the values of $p$, $q$ and a set of $\{a_m\}$.
The spacetime interpretation of the lattice is given 
through the linear mapping 
(\ref{linear mapping for N=3}) in which the values 
of $(\theta_1,\theta_2)$ and $\gamma_m$ are determined by 
(\ref{values of theta}) and (\ref{gamma wrt pqa}). 
Although there are infinitely many theories, we can again classify them
by the number of the remaining supersymmetries. 
As we shall see, they give different lattice formulations 
of what in the continuum becomes
two-dimensional $\cN=(4,4)$ supersymmetric gauge theory.

\vspace{3mm}
\noindent
[2-1] $\bfe_3=-\bfe_1$ (or $-\bfe_2$)

This is the case of $(p,q)=(-1,0)$ (or $(0,-1)$) and we assume that 
all $a_m$ have finite values. 
{}From (\ref{values of theta}), we see $(\theta_2,\theta_3)=(\pi/2,-\pi)$ 
(or $(\pi/2,3\pi/2)$), then the lattice is a square lattice. 
The values of $\gamma_m$ are given by 
\begin{align}
 (\gamma_1^2,\gamma_2^2,\gamma_3^2)&= 
\Bigl(\frac{a_3^2 l^2}{a_1^2+a_3^2},\, l^2,\, 
 \frac{a_1^2 l^2}{a_1^2+a_3^2}\Bigr). \quad 
\Biggl({\rm or}\ 
\Bigl(l^2,\, \frac{a_3^2 l^2}{a_2^2+a_3^2},\, 
 \frac{a_2^2 l^2}{a_2^2+a_3^2}\Bigr)
\Biggr)
\end{align}
Although the action is different, 
the field configuration of this theory is the same with [1-2] 
and thus there are two conserved supercharges.  
(See (b) of Fig. \ref{8SUSY_d=D=2 Q=2}.)

The roles of the fields in the continuum limit are slightly
non-trivial. To see this, it is again useful to look at the kinetic terms
of $z_m$ in the continuum limit, 
\begin{align}
 -l^2 S_m(\bfx)\del^2 S_m(\bfx)
 -T_m(\bfx)\Bigl[l^2 \del^2 \delta_{mn}
-\left(\gamma_m\hmu_m\cdot\vec\del\right)
 \left(\gamma_n\hmu_n\cdot\vec\del\right)\Bigr]T_n(\bfx) , 
\label{kinetic terms for D=3}
\end{align}
where $l,m,n=1,2,3$, and $S_m$ and $T_m$ are the real and imaginary 
components of $z_m$, respectively. 
We then define an orthogonal matrix $P_{mn}$ \cite{Kaplan:2005ta}, 
\begin{equation}
 \sum_{n=1}^3 P_{mn} \gamma_n \hmu_n = 
\begin{cases}
 \left(\begin{matrix} l \\0 \end{matrix}\right), & (m=1) \\
 \left(\begin{matrix} 0\\ l \end{matrix}\right), & (m=2) \\
 0, & (m=3)
\end{cases} 
\end{equation}
with the help of which we can rewrite $T_m$ as 
\begin{equation}
 T_m \equiv \sum_{\mu=1,2} \frac{1}{l} \left(P_{m\mu} A_\mu\right) 
+ \frac{1}{l} P_{m3} S_4. 
\end{equation}
Finally (\ref{kinetic terms for D=3}) can be rewritten as 
\begin{equation}
 -\sum_{a=1}^4 S_a(\bfx)\del^2 S_a(\bfx)
 -\sum_{\mu,\nu=1}^2 A_\mu(\bfx)\Bigl[\del^2 \delta_{\mu\nu}
-\del_\mu \del_{\nu}\Bigr]A_\nu(\bfx) ,
\end{equation}
where we have also rescaled $S_m\to\frac{1}{l}S_m$. 
This is nothing but the canonical kinetic terms for scalar 
bosons and a gauge vector. 
As a result, the continuum limit of this theory has four real scalar
fields and one vector field. 
One can also show that the kinetic terms of fermions and the interaction
terms become that of two-dimensional $\cN=(4,4)$ supersymmetric gauge
theory in the naive continuum limit.

\begin{figure}[t]
  \begin{center}
   \parbox{.4\linewidth}{\begin{center}
    \includegraphics[scale=1]{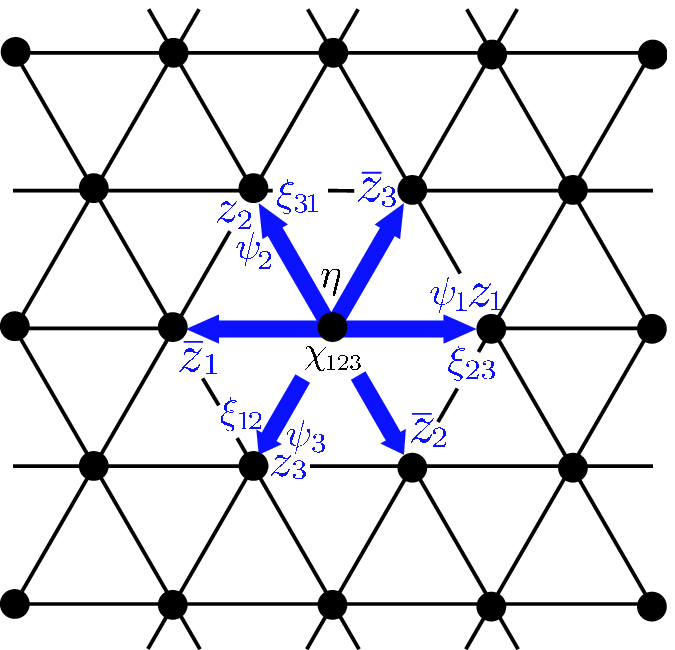}\\
    (a) $\bfe_3=-\bfe_1-\bfe_2$\end{center}}
\hspace{1cm}
    \parbox{.4\linewidth}{\begin{center}
    \includegraphics[scale=1]{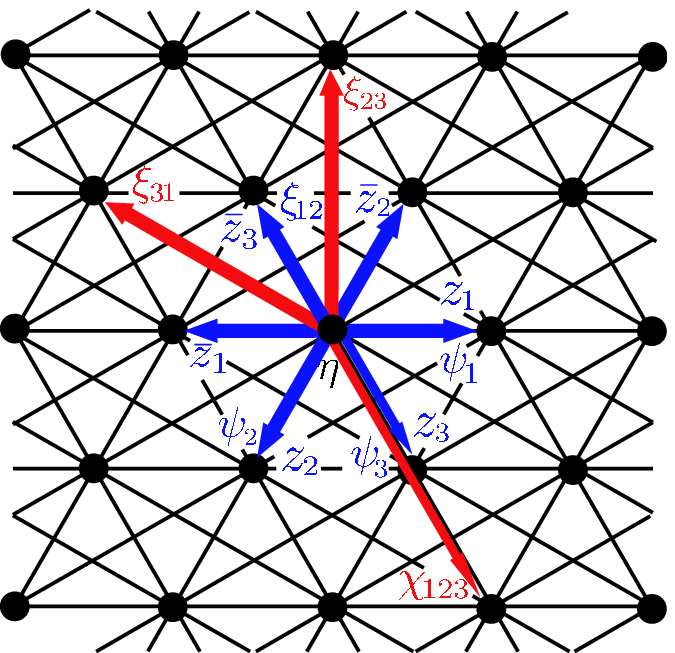}\\
    (b) $\bfe_3=\bfe_1+\bfe_2$\end{center}}
  \end{center}
  \caption{The lattice structure and the field configuration of the
 formulations [2-2] and [2-3]. 
 The left figure (a) expresses the lattice structure and the field
 configurations for the case of $p=q=-1$ where 
 there are two remaining supersymmetries. 
   The right figure (b) expresses the case of $p=q=1$, where there is one
 remaining supersymmetry. 
 For simplicity, we have chosen $a_1=a_2=a_3$ for both the cases. 
}
  \label{8SUSY_d=2_D=3}
\end{figure}

\vspace{3mm}
\noindent
[2-2] $\bfe_1+\bfe_2+\bfe_3=0$

This is the case of $p=q=-1$ and we again assume all $a_m$ are finite. 
Since $p$ and $q$ are negative, $\theta_2$ and $\theta_3$ are 
in the region (\ref{two SUSY case}), 
thus the lattice is triangular in general. 
Looking at Table \ref{integer charge table 8 SUSY}, 
$\chi_{123}$ turns out to be a singlet, 
so that this lattice theory possesses two
conserved supercharges. 
Furthermore, in this formulation all fermions live on links of the
triangles and there are no ``diagonal'' links. 
In fact, $\eta$ and $\chi_{123}$ live on sites, and
$z_m$ and $\phi_m$ $(m=1,2,3)$ are link variables 
in the directions of
$\hmu_m$ given by (\ref{values of theta}). 
Also $\xi_{23}$, $\xi_{31}$ and $\xi_{12}$ live on the links, 
in the directions of $-\hmu_1$, $-\hmu_2$ and $-\hmu_3$, respectively.
(See (a) of Fig. \ref{8SUSY_d=2_D=3}.) 

Using the same logic as in [2-1], we can show that the bosonic fields 
behave as a gauge field and six scalar fields in the naive continuum
limit and the continuum theory is again expected to be two-dimensional 
$\cN=(4,4)$ supersymmetric gauge theory.

\vspace{3mm}
\noindent
[2-3] $\bfe_3 \notin \{0,\, -\bfe_1,\,-\bfe_2,\, -\bfe_1-\bfe_2\} $

For any other rational values of $p$ and $q$ except for the
special cases $(p,q)=(0,0)$, $(-1,0)$, $(0,-1)$ and $(-1,-1)$, 
$\eta$ is the only scalar fermion. The orbifolded
lattice theories will then have only one preserved supersymmetry. 
The lattice structure is less symmetric than the above case 
and there are several ``diagonal'' link variables in general. 
As a simple example, consider the case of $p=q=1$ 
($\bfe_3 = \bfe_1 + \bfe_2$). 
(See (b) of Fig. \ref{8SUSY_d=2_D=3}.)
In this case, Lorentz invariance in the continuum forces 
the set $\{\hmu_m\}$ to be in the region (\ref{one SUSY case}). 
Now, $\xi_{12}$ and $\xi_{31}$ live on the diagonal links 
$(\bfk,\bfk-\bfe_1-\bfe_2)$ and $(\bfk,\bfk-\bfe_3-\bfe_1)$,
respectively,  
and $\chi_{123}$ lives on the links $(\bfk,\bfk+2\bfe_1)$. 
For the other fields, $\eta$ sits on sites, 
while $z_m$, $\bar{z}_m$, $\psi_m$ and $\xi_{23}$ live on the ordinary links. 
Using the same argument as above, we can show that 
the real components of $z_m$ and a linear combination of the
imaginary components of $z_m$ become real scalar fields, 
and the other linearly independent components of the imaginary parts of
$z_m$ form a vector field in the continuum limit. 
Thus the naive continuum limit is again two-dimensional 
$\cN=(4,4)$ supersymmetric gauge theory. 

\subsubsection{Relation between $N_{\rm shift}=2$ and $N_{\rm shift}=3$}

In this short subsection, we will mention an interesting connection between 
the cases of $N_{\rm shift}=2$ and $N_{\rm shift}=3$ discussed 
in sections \ref{Ns=2} and \ref{Ns=3}, respectively.

In the case of $N_{\rm shift}=3$, the space-time lattice is uniquely
determined by (\ref{values of theta}) and (\ref{gamma wrt pqa}) 
for given values of $(p,q)$ and $\{a_m\}$, 
where we assumed that the values of $a_m$ are all finite.  
However, we can easily see that 
these formulae can be applied even 
for the case where one of the $a_m$ go to infinity. 
Indeed, if we take the limit of $a_3\to\infty$ with $p,q\ne0$, 
we obtain $\tan\theta_2=\pm \infty$ and 
$\tan\theta_3=\mp\frac{qa_2}{p a_1}$, where the upper sign is for $p,q>0$ 
and the lower sign is for $p,q<0$. 
This means that the two vectors $\hmu_1$ and
$\hmu_2$ are orthogonal to each other and the direction of $\hmu_3$
is the same as $\bfe_3$ given by the linear relation 
(\ref{linear relation}).
Furthermore, one can explicitly show 
that the combinations $\gamma_m a_m \hmu_m$
satisfy the relation (\ref{linear condition after mapping}). 
Then, recalling the definition of the shift (\ref{shift of zm}),  
we conclude that the case of $N_{\rm shift}=3$
contains the case of $N_{\rm shift}=2$ as a special case. 
This result is true also for the cases of $p=0$ or $q=0$. 
In fact, 
(\ref{values of theta}) and (\ref{gamma wrt pqa}) can be applied 
even then
and we can safely take the limit of $a_3\to\infty$ for 
the case of $p=0$ and $q\ne 0$, and $a_1\to\infty$ for 
the case of $p\ne 0$ and $q= 0$. 
When $p=q=0$, the theory becomes automatically [1-1]
for any value of $a_3$. 

As could have been expected intuitively, we thus find that all 
2-shift solutions are just special cases of the general 3-shift
solutions.

\section{Conclusions}

Following refs.~\cite{Cohen:2003xe,Cohen:2003qw},  
we have considered the dimensionally reduced theories of four-dimensional
$\cN=1$ SYM theory and six-dimensional $\cN=1$ SYM theory, and viewed
them as ``mother theories'' for orbifolded lattice field theories. 
We have given what we believe is a complete classification 
of all possible lattice gauge theories in dimensions larger than or
equal to two that can be
constructed from these mother theories by the orbifolding procedure 
given in \cite{Cohen:2003xe,Cohen:2003qw}. 
We have imposed on the lattice theories that they 
have at least one preserved scalar supercharge, 
and that they become Lorentz invariant in the naive continuum limit. 

Starting with the mother theory with four supercharges 
we have found that there is only one lattice formulation
possible by this route. Its continuum
limit is two-dimensional $\cN=(2,2)$ supersymmetric gauge theory. 
This formulation is identical to what was given in \cite{Cohen:2003xe}. 
We have thus shown that this formulation is unique.

On the other hand, starting with a mother theory with eight
supercharges, there are many more possibilities. 
One can construct both three-dimensional 
and two-dimensional lattice theories in this case. 
We have found that the three-dimensional theory is again unique, 
and it coincides with the one given in \cite{Cohen:2003qw}. 
For the two-dimensional theories, however,  
one can construct infinitely many lattice formulations labelled by 
two rational numbers $p$ and $q$. 
We have shown that they can be classified into seven 
categories by the number of remaining supersymmetries and 
the structure of the lattice. 
Five of these have two preserved scalar supercharges; 
the others have one. 
In the naive continuum limit, these formulations yield 
the same theory:
two-dimensional $\cN=(4,4)$ supersymmetric gauge theory. 
The five formulations with two supersymmetries, 
[1-1], [1-2], [1-3], [2-1] and [2-2] are in a sense cousins.
In fact, we can reach these formulations by tuning one of the
fermions $\psi_m$, $\xi_{mn}$ and $\chi_{123}$ to be a singlet under the
$U(1)$ symmetries.
In these formulations, the space-time lattices are highly symmetric and 
form simple tilings of the two-dimensional plane. 
On the other hand, the lattice structures of the theories with 
one scalar supercharge, [1-4] and [2-3], are less symmetric. 
Therefore, even if the continuum limit is the same at tree level, 
one might prefer those lattice formulations 
that are closer to continuum Lorentz invariance already at finite
lattice spacings.  
In this paper we have insisted on Lorentz invariance in the continuum 
limit. Since the lattice theories in question have at least one
preserved supercharge at all lattice spacings
it could be interesting to consider what
types of continuum theories might emerge if one relaxes this condition.

\vspace{0.5cm}
\noindent
{\sc Acknowledgement:}~ 
We thank 
S.~Hirano,
I.~Kanamori,
K.~Ohta,  
H.~Suzuki, 
and T.~Takimi
for useful discussions.
S.M. also acknowledges support from 
JSPS Postdoctoral Fellowship for Research Abroad.

\appendix
\section{Useful Formulae}

In this appendix, we briefly review the orbifold projection 
and summarize some useful formulae. 
Let us consider a matrix theory (mother theory) and 
an adjoint field $\Phi$ in it, transforming as 
$\Phi \to g^{-1} \Phi g$ under the $U(kN^n)$ ``gauge'' symmetry. 
We also assume that this theory is invariant under 
a ``global'' symmetry $R$. 
In our case, $R=SO(4)\times U(1)$ for (\ref{Smother0})
and $R=SO(6)\times U(1)$ for (\ref{mather action for 8 SUSY}). 
Suppose that $R$ contains $U(1)$ subgroups, $U(1)^n$, 
and $\Phi$ carries integer charges $(q_1,\cdots,q_n)$ $(q_a\in \Z)$. 

Under these assumptions, we consider a $Z_N^n$ symmetry generated by 
\begin{equation}
 \gamma_a: \Phi \to \omega^{-q_a} \Omega_a^{-1} \Phi \Omega_a, 
 \qquad (a=1,\cdots,n) 
\end{equation}
where $\omega=e^{2\pi i/N}$ and 
\begin{equation}
 \Omega_a \equiv 1_k \otimes 
 \underbrace{1_N \otimes 1_N}_{a-1} \otimes U \otimes 
 \underbrace{1_N \otimes 1_N}_{n-a}, \qquad 
 U\equiv {\rm diag}(\omega^1,\cdots,\omega^N). 
\end{equation}
Using $\gamma_a$, we can define a projection operator,  
\begin{equation}
 P\equiv \frac{1}{N^n} \sum_{k_1,\cdots,k_n=1}^N 
 \gamma_1^{k_1}\cdots \gamma_n^{k_n}. 
\end{equation}
Using the relation 
\begin{equation}
 U^{-1}V U = \omega V, \qquad 
 V \equiv\left(\begin{matrix}
  0 & & 1 &  \\
    & & \ddots & \ddots \\
    &       &  &   0    & 1 \\
  1 &       &  &        & 0  
               \end{matrix}\right), 
\label{projected Phi}
\end{equation}
we can easily show that the projected matrix by $P$ can be expressed
as 
\begin{align}
 \Phi &=\sum_{\bfm\in \Z_N^n}
\tilde{\Phi}(\bfm)\otimes U^{m_1}V^{q_1}\otimes\cdots\otimes
 U^{m_n}V^{q_n},
\end{align}
or equivalently, 
\begin{align}
 \Phi &= \sum_{\bfk \in \Z_N^n} {\Phi}(\bfk)\otimes E_{\fatk,\fatk+\bfq}, 
\end{align}
where $\tilde\Phi(\bfm)$ and $\Phi(\bfk)$ are $k\times k$ matrices and 
we have defined 
\be
E_{\fatk,{\mathbf l}} ~=~ E_{k_1,l_1}\otimes \cdots \otimes E_{k_n,l_n},
\label{basis matrices}
\ee
with 
\begin{equation}
 (E_{l,m})_{ij} \equiv \delta_{li}\delta_{mj}.
\end{equation}
The orbifold projection is defined by restricting fields in the mother
theory to those which are invariant by the operation of $P$. 
Then, by construction, the orbifolded action is 
obtained by substituting (\ref{projected Phi}) into the action
of the mother theory. 
The orbifolded actions (\ref{Sorb0}) and (\ref{orbifold action 8 SUSY})
are obtained by this procedure. 
In calculating the orbifolded action, the relation, 
\be
E_{\bfk,\bfl}E_{\bfm,\bfn} = \delta_{\bfl\bfm}E_{\bfk,\bfn}.
\label{orthogonality relation 2}
\ee
is quite useful, which directly comes from the relation, 
\be
E_{i,j}E_{k,l} ~=~ \delta_{jk}E_{i,l}. 
\label{orthogonality relation}
\ee

\section{Supersymmetry Transformations}
In this appendix, we derive the explicit supersymmetry transformations 
of lattice theories constructed in this paper. 
We concentrate on the theories that are derived from the mother
theory with eight supercharges, since it is this case which leads
to new lattice
formulations. The derivation for the other case (with four supercharges)
is completely parallel. 

Our treatment builds heavily on the very clear discussion in ref.
\cite{Cohen:2003qw}. We thus 
start with the supersymmetry transformations of the mother theory,  
and rewrite 
the action (\ref{mather action for 8 SUSY}) as 
\begin{align}
 S_m &= \frac{1}{g^2}\Tr_N \biggl(
\frac{1}{4} v_{\alpha\beta}^2 -\frac{i}{2}{\rm tr}_2 
\Bigl(\tau_2 \Psi^T C \bar\Sigma_\alpha [v_\alpha, \Psi]\Bigr)
\biggr),
\label{8 SUSY mother 2}
\end{align}
where $v_{\alpha\beta}\equiv i[v_\alpha,v_\beta]$, 
${\rm tr}_2$ denotes the trace over 
$2\times 2$ matrix, and $\Psi$ is defined by 
\begin{equation}
 \Psi \equiv \Bigl(\psi, C \bpsi^T\Bigr), 
\end{equation}
with a ``charge conjugation matrix'' $C$ satisfying 
\begin{equation}
 C^\dagger \bar\Sigma_m C = \bar\Sigma_m^T,\qquad  
 C=C^\dagger = C^{-1} = - C^T. 
\end{equation}
Note that, in this notation, the configuration (\ref{fermion configuration}) 
corresponds to 
\begin{equation}
 \Psi = \left(\begin{matrix}
 \eta & -i \chi_{123} \\
 \xi_{23} & -i \psi_1 \\
 \xi_{31} & -i \psi_2 \\
 \xi_{12} & -i \psi_3
\end{matrix}\right), 
\end{equation}
where we have used the representation, 
\begin{equation}
 C=\left(\begin{matrix}
0 & -i & 0 & 0 \\
i & 0 & 0 & 0 \\
0 & 0 & 0 & i \\
0 & 0 & -i & 0 \\
\end{matrix}\right). 
\end{equation}
One can easily check that (\ref{8 SUSY mother 2}) coincides with 
(\ref{mather action for 8 SUSY}). 
In this notation, the supersymmetric transformation can be expressed
compactly as 
\begin{align}
 \delta v_\alpha&= {\rm tr}_2 \left(\tau_2 \kappa^T C \bar\Sigma_\alpha
 \Psi \right) 
 \nn \\
 \delta \Psi &= -i v_{\alpha\beta}\Sigma_{\alpha\beta}\kappa, 
\label{SUSY trans.}
\end{align}
where $\kappa$ is a constant Grassmann parameter with the form of
a $4\times 2$ matrix.

We recall that the remaining supercharges on the lattice should correspond
to fermions that have zero $U(1)$ charges. Furthermore, the supersymmetry
parameter matrix $\kappa$ has the same structure as $\Psi$. 
We see that the supersymmetry transformations of the orbifolded theory 
therefore
can be obtained by restricting $\kappa$ correspondingly, 
followed by the orbifolding projection.  
In the following, we derive those transformations that leave invariant
the actions of the two-dimensional theories 
[1-2] and [2-1] ($\bfe_1+\bfe_3=0$), 
and [1-3] and [2-1] ($\bfe_1+\bfe_2+\bfe_3=0$) 
discussed in the section 3. 
For the supersymmetry transformations of 
the three-dimensional theory and
the two-dimensional theory [1-1], see ref.~\cite{Cohen:2003qw}. 
Those of the theories [1-4] and [2-3] are essentially the same as
the three-dimensional theory, and we do not display them explicitly here.

\subsection{Supersymmetry transformations of the models 1-2 and 2-1}

In these cases, the $U(1)$ charges of 
$\eta$ and $\xi_{31}$ have $U(1)$ are zero. We thus fix the
supersymmetry parameter matrix as 
\begin{equation}
\kappa = \left(\begin{matrix}
 \kappa_1 & 0 \\
 0 & 0 \\
 \kappa_2 & 0 \\
 0 & 0 
\end{matrix}\right). 
\end{equation}
Substituting this into (\ref{SUSY trans.}), we obtain 
\begin{alignat}{2}
 \delta z_1&= -2i \kappa_1 \psi_1, &\quad \delta\bz_1&= 2i\kappa_2 \psi_3, \nn
 \\
 \delta z_2&= -2i \kappa_1 \psi_2+2i \kappa_2 \chi_{123}, 
  &\quad \delta\bz_2&= 0, \nn \\
 \label{SUSY of orbifolded 1-2}
 \delta z_3&= -2i \kappa_1 \psi_3, &\quad \delta\bz_3&= -2i\kappa_2 \psi_1,
 \\
 \delta \eta &= \frac{i}{4}\sum_{m=1}^3[z_m,\bz_m] \kappa_1 
  +\frac{i}{2}[z_1,z_3]\kappa_2, &\quad   
  \delta \xi_{31} &= \frac{i}{2}[\bz_3,\bz_1]\kappa_1 
  - \frac{i}{4}\left([z_1,\bz_1]-[z_2,\bz_2]+[z_3,\bz_3]\right)\kappa_2
 \nn \\ 
 \delta \xi_{23} &= \frac{i}{2}[\bz_2,\bz_3]\kappa_1 
  + \frac{i}{2}[z_1,\bz_2]\kappa_2, &\quad 
 \delta \xi_{12} &= \frac{i}{2}[\bz_1,\bz_2]\kappa_1 
  + \frac{i}{2}[z_3,\bz_2]\kappa_2, \nn \\ 
\delta \psi_m &=0, &  \delta \chi_{123}&=0. \nn
\end{alignat}
Correspondingly, we can define two supercharges $Q_1$ and $Q_2$ as 
\begin{equation}
 \delta = 2i\kappa_1 Q_1 + 2i\kappa_2 Q_2, 
\label{Q1 and Q2}
\end{equation}
which satisfy $Q_1^2=Q_2^2=0$ on-shell. 
In order to make the nilpotency satisfy off-shell, we introduce an
auxiliary field $d$ and modify the transformations
of $\eta$ and $\xi_{31}$ as 
\begin{align}
 \delta \eta &= 
 \Bigl(\frac{i}{4}\sum_{m=1}^3[z_m,\bz_m]-2i d \Bigr) \kappa_1 
  +\frac{i}{2}[z_1,z_3]\kappa_2, \nn \\
 \delta \xi_{31} &= \frac{i}{2}[\bz_3,\bz_1]\kappa_1 
+\Bigl(
- \frac{i}{4}\left([z_1,\bz_1]-[z_2,\bz_2]+[z_3,\bz_3]\right)
+2id
\Bigr)  \kappa_2, 
\end{align}
where the transformation of $d$ is 
\begin{equation}
 \delta d = -\frac{i}{4}\sum_{m=1}^3[\psi_m,\bz_m] \kappa_1 
  +\frac{i}{4}\Bigl([z_1,\psi_3]-[z_3,\psi_1]-[\chi_{123},\bz_2]\Bigr)\kappa_2.
\end{equation}
Then $Q_1^2=Q_2^2=\{Q_1,Q_2\}=0$. 
The transformations (\ref{SUSY of orbifolded 1-2}) are those of the
mother theory. 
The corresponding supersymmetry transformations of the lattice theory are obtained by
substituting the expansion (\ref{expansion}) into 
(\ref{SUSY of orbifolded 1-2}). It is tedious but straightforward to check
that they indeed leave the lattice theory invariant.

\subsection{Supersymmetry transformations of the models 1-3 and 2-2}

In these cases,
the $U(1)$ charges of $\eta$ and $\chi_{123}$ are zero, and we therefore 
fix $\kappa$ to be 
\begin{equation}
\kappa = \left(\begin{matrix}
 \kappa_1 & -i\kappa_2 \\
 0 & 0 \\
 0 & 0 \\
 0 & 0 
\end{matrix}\right). 
\end{equation} 
Substituting this into (\ref{SUSY trans.}), we obtain 
\begin{alignat}{2}
 \delta z_l &= -2i \kappa_1 \psi_l +i\kappa_2
 \sum_{m,n=1}^3\e_{lmn}\xi_{mn},  &\quad 
 \delta \bz_m &= 0, \nn \\
 \label{SUSY for 1-3}
 \delta \eta &= \frac{i}{4}\sum_{m=1}^3[z_m,\bz_m]\kappa_1, &\quad 
 \delta \chi_{123} &= \frac{i}{4}\sum_{m=1}^3[z_m,\bz_m]\kappa_2, \\
 \delta \psi_l &=
 \frac{i}{4}\sum_{m,n=1}^3\e_{lmn}[\bz_m,\bz_n]\kappa_2,  
  &\quad 
 \delta \xi_{mn} &= \frac{i}{2}[\bz_m,\bz_n] \kappa_1. \nn
\end{alignat}
Again, we can define two supercharges $Q_1$ and $Q_2$ by 
(\ref{Q1 and Q2}), and we can again make them nilpotent off-chell by 
introducing an auxiliary field $d$ in the transformations of $\eta$ and
$\chi_{123}$:
\begin{align}
 \delta \eta &= \Bigl(\frac{i}{4}\sum_{m=1}^3[z_m,\bz_m]
 -2id\Bigr)\kappa_1, \nn \\
 \delta \chi_{123} &= \Bigl(
\frac{i}{4}\sum_{m=1}^3[z_m,\bz_m]-2id
\Bigr)\kappa_2,
\end{align}
where the transformation of $d$ is defined as 
\begin{equation}
 \delta d = -\frac{i}{4}\sum_{m=1}^3[\psi_m,\bz_m] \kappa_1
  +\frac{i}{8}\sum_{l,m,n=1}^3\e_{lmn}[\xi_{mn},\bz_l]\kappa_2. 
\end{equation}
One can explicitly check that $Q_1$ and $Q_2$ now satisfy 
$Q_1^2=Q_2^2=\{Q_1,Q_2\}=0$. 
Again, the supersymmetry transformations for the lattice theory are
obtained by substituting the expansion (\ref{expansion}) into 
(\ref{SUSY for 1-3}).

\bibliographystyle{JHEP}
\bibliography{refs}

\providecommand{\href}[2]{#2}\begingroup\raggedright\begin{thebibliography}{10}

\bibitem{Kaplan:2002wv}
D.~B. Kaplan, E.~Katz and M.~Unsal, {\it Supersymmetry on a spatial lattice},
  {\em JHEP} {\bf 05} (2003) 037
  [\href{http://arXiv.org/abs/hep-lat/0206019}{{\tt hep-lat/0206019}}].

\bibitem{Cohen:2003xe}
A.~G. Cohen, D.~B. Kaplan, E.~Katz and M.~Unsal, {\it Supersymmetry on a
  Euclidean spacetime lattice. I: A target theory with four supercharges},
  {\em JHEP} {\bf 08} (2003) 024
  [\href{http://arXiv.org/abs/hep-lat/0302017}{{\tt hep-lat/0302017}}].

\bibitem{Cohen:2003qw}
A.~G. Cohen, D.~B. Kaplan, E.~Katz and M.~Unsal, {\it Supersymmetry on a
  Euclidean spacetime lattice. II: Target theories with eight supercharges},
  {\em JHEP} {\bf 12} (2003) 031
  [\href{http://arXiv.org/abs/hep-lat/0307012}{{\tt hep-lat/0307012}}].

\bibitem{Kaplan:2005ta}
D.~B. Kaplan and M.~Unsal, {\it A Euclidean lattice construction of
  supersymmetric Yang- Mills theories with sixteen supercharges},  {\em JHEP}
  {\bf 09} (2005) 042 [\href{http://arXiv.org/abs/hep-lat/0503039}{{\tt
  hep-lat/0503039}}].

\bibitem{Endres:2006ic}
M.~G. Endres and D.~B. Kaplan, {\it Lattice formulation of (2,2) supersymmetric
  gauge theories with matter fields},  {\em JHEP} {\bf 10} (2006) 076
  [\href{http://arXiv.org/abs/hep-lat/0604012}{{\tt hep-lat/0604012}}].

\bibitem{Catterall:2001fr}
S.~Catterall and S.~Karamov, {\it Exact lattice supersymmetry: the
  two-dimensional N = 2 Wess-Zumino model},  {\em Phys. Rev.} {\bf D65} (2002)
  094501 [\href{http://arXiv.org/abs/hep-lat/0108024}{{\tt hep-lat/0108024}}].

\bibitem{Catterall:2001wx}
S.~Catterall and S.~Karamov, {\it A two-dimensional lattice model with exact
  supersymmetry},  {\em Nucl. Phys. Proc. Suppl.} {\bf 106} (2002) 935--937
  [\href{http://arXiv.org/abs/hep-lat/0110071}{{\tt hep-lat/0110071}}].

\bibitem{Catterall:2003uf}
S.~Catterall and S.~Ghadab, {\it Lattice sigma models with exact
  supersymmetry},  {\em JHEP} {\bf 05} (2004) 044
  [\href{http://arXiv.org/abs/hep-lat/0311042}{{\tt hep-lat/0311042}}].

\bibitem{Catterall:2003wd}
S.~Catterall, {\it Lattice supersymmetry and topological field theory},  {\em
  JHEP} {\bf 05} (2003) 038 [\href{http://arXiv.org/abs/hep-lat/0301028}{{\tt
  hep-lat/0301028}}].

\bibitem{Catterall:2005fd}
S.~Catterall, {\it Lattice formulation of N = 4 super Yang-Mills theory},  {\em
  JHEP} {\bf 06} (2005) 027 [\href{http://arXiv.org/abs/hep-lat/0503036}{{\tt
  hep-lat/0503036}}].

\bibitem{Catterall:2006jw}
S.~Catterall, {\it Simulations of N = 2 super Yang-Mills theory in two
  dimensions},  {\em JHEP} {\bf 03} (2006) 032
  [\href{http://arXiv.org/abs/hep-lat/0602004}{{\tt hep-lat/0602004}}].

\bibitem{D'Adda:2004jb}
A.~D'Adda, I.~Kanamori, N.~Kawamoto and K.~Nagata, {\it Twisted superspace on a
  lattice},  {\em Nucl. Phys.} {\bf B707} (2005) 100--144
  [\href{http://arXiv.org/abs/hep-lat/0406029}{{\tt hep-lat/0406029}}].

\bibitem{D'Adda:2005zk}
A.~D'Adda, I.~Kanamori, N.~Kawamoto and K.~Nagata, {\it Exact extended
  supersymmetry on a lattice: Twisted N = 2 super Yang-Mills in two
  dimensions},  {\em Phys. Lett.} {\bf B633} (2006) 645--652
  [\href{http://arXiv.org/abs/hep-lat/0507029}{{\tt hep-lat/0507029}}].

\bibitem{Bruckmann:2006ub}
F.~Bruckmann and M.~de~Kok, {\it Noncommutativity approach to supersymmetry on
  the lattice: SUSY quantum mechanics and an inconsistency},  {\em Phys. Rev.}
  {\bf D73} (2006) 074511 [\href{http://arXiv.org/abs/hep-lat/0603003}{{\tt
  hep-lat/0603003}}].

\bibitem{Bruckmann:2006kb}
F.~Bruckmann, S.~Catterall and M.~de~Kok, {\it A critique of the link approach
  to exact lattice supersymmetry},  {\em Phys. Rev.} {\bf D75} (2007) 045016
  [\href{http://arXiv.org/abs/hep-lat/0611001}{{\tt hep-lat/0611001}}].

\bibitem{Sugino:2003yb}
F.~Sugino, {\it A lattice formulation of super Yang-Mills theories with exact
  supersymmetry},  {\em JHEP} {\bf 01} (2004) 015
  [\href{http://arXiv.org/abs/hep-lat/0311021}{{\tt hep-lat/0311021}}].

\bibitem{Sugino:2004qd}
F.~Sugino, {\it Super Yang-Mills theories on the two-dimensional lattice with
  exact supersymmetry},  {\em JHEP} {\bf 03} (2004) 067
  [\href{http://arXiv.org/abs/hep-lat/0401017}{{\tt hep-lat/0401017}}].

\bibitem{Sugino:2004uv}
F.~Sugino, {\it Various super Yang-Mills theories with exact supersymmetry on
  the lattice},  {\em JHEP} {\bf 01} (2005) 016
  [\href{http://arXiv.org/abs/hep-lat/0410035}{{\tt hep-lat/0410035}}].

\bibitem{Sugino:2006uf}
F.~Sugino, {\it Two-dimensional compact N = (2,2) lattice super Yang-Mills
  theory with exact supersymmetry},  {\em Phys. Lett.} {\bf B635} (2006)
  218--224 [\href{http://arXiv.org/abs/hep-lat/0601024}{{\tt
  hep-lat/0601024}}].

\bibitem{Unsal:2005yh}
M.~Unsal, {\it Compact gauge fields for supersymmetric lattices},  {\em JHEP}
  {\bf 11} (2005) 013 [\href{http://arXiv.org/abs/hep-lat/0504016}{{\tt
  hep-lat/0504016}}].

\bibitem{Unsal:2006qp}
M.~Unsal, {\it Twisted supersymmetric gauge theories and orbifold lattices},
  {\em JHEP} {\bf 10} (2006) 089
  [\href{http://arXiv.org/abs/hep-th/0603046}{{\tt hep-th/0603046}}].

\bibitem{Arkani-Hamed:2001ca}
N.~Arkani-Hamed, A.~G. Cohen and H.~Georgi, {\it (De)constructing dimensions},
  {\em Phys. Rev. Lett.} {\bf 86} (2001) 4757--4761
  [\href{http://arXiv.org/abs/hep-th/0104005}{{\tt hep-th/0104005}}].

\bibitem{Giedt:2003xr}
J.~Giedt, E.~Poppitz and M.~Rozali, {\it Deconstruction, lattice supersymmetry,
  anomalies and branes},  {\em JHEP} {\bf 03} (2003) 035
  [\href{http://arXiv.org/abs/hep-th/0301048}{{\tt hep-th/0301048}}].

\bibitem{Giedt:2003ve}
J.~Giedt, {\it Non-positive fermion determinants in lattice supersymmetry},
  {\em Nucl. Phys.} {\bf B668} (2003) 138--150
  [\href{http://arXiv.org/abs/hep-lat/0304006}{{\tt hep-lat/0304006}}].

\bibitem{Onogi:2005cz}
T.~Onogi and T.~Takimi, {\it Perturbative study of the supersymmetric lattice
  theory from matrix model},  {\em Phys. Rev.} {\bf D72} (2005) 074504
  [\href{http://arXiv.org/abs/hep-lat/0506014}{{\tt hep-lat/0506014}}].

\bibitem{Ohta:2006qz}
K.~Ohta and T.~Takimi, {\it Lattice formulation of two dimensional topological
  field theory},  \href{http://arXiv.org/abs/hep-lat/0611011}{{\tt
  hep-lat/0611011}}.

\bibitem{Fukaya:2006mg}
H.~Fukaya, I.~Kanamori, H.~Suzuki, M.~Hayakawa and T.~Takimi, {\it Note on
  massless bosonic states in two-dimensional field theories},  {\em Prog.
  Theor. Phys.} {\bf 116} (2007) 1117--1129
  [\href{http://arXiv.org/abs/hep-th/0609049}{{\tt hep-th/0609049}}].

\bibitem{Giedt:2006pd}
J.~Giedt, {\it Deconstruction and other approaches to supersymmetric lattice
  field theories},  {\em Int. J. Mod. Phys.} {\bf A21} (2006) 3039--3094
  [\href{http://arXiv.org/abs/hep-lat/0602007}{{\tt hep-lat/0602007}}].

\bibitem{Rabin:1981qj}
J.~M. Rabin, {\it HOMOLOGY THEORY OF LATTICE FERMION DOUBLING},  {\em Nucl.
  Phys.} {\bf B201} (1982) 315.

\bibitem{Becher:1982ud}
P.~Becher and H.~Joos, {\it The Dirac-Kahler Equation and Fermions on the
  Lattice},  {\em Zeit. Phys.} {\bf C15} (1982) 343.

\bibitem{Suzuki:2005dx}
H.~Suzuki and Y.~Taniguchi, {\it Two-dimensional N = (2,2) super Yang-Mills
  theory on the lattice via dimensional reduction},  {\em JHEP} {\bf 10} (2005)
  082 [\href{http://arXiv.org/abs/hep-lat/0507019}{{\tt hep-lat/0507019}}].

\bibitem{DM}
P.~H. Damgaard and S.~Matsuura, {\it in preparation}.

\bibitem{Eguchi:1982nm}
T.~Eguchi and H.~Kawai, {\it Reduction of Dynamical Degrees of Freedom in the
  Large N Gauge Theory},  {\em Phys. Rev. Lett.} {\bf 48} (1982) 1063.

\bibitem{Orland:1983gt}
P.~Orland, {\it VOLUME REDUCTION OF LATTICE GAUGE SYSTEMS AT FINITE N},  {\em
  Phys. Lett.} {\bf B134} (1984) 95.

\end{thebibliography}\endgroup

\end{document}